\documentclass[a4paper,fleqn,usenatbib]{mnras}

\usepackage{newtxtext,newtxmath}
\usepackage[T1]{fontenc}
\usepackage{ae,aecompl}

\usepackage{graphicx}
\usepackage{amsmath}
\usepackage{amssymb}
\usepackage{bm}
\usepackage{natbib}

\usepackage[caption = false]{subfig} 

\usepackage{trace}
\usepackage{listings}
\usepackage{alltt}
\usepackage[normalem]{ulem} 
\usepackage{tablefootnote}
\usepackage{threeparttable} 
\usepackage{cprotect}

\newcommand{\rhalf}{r_{\rm{half}}}

\newcommand{\gb}{$g$}
\newcommand{\rband}{$r$}
\newcommand{\zb}{$z$}

\newcommand{\grcolor}{$g - r$}
\newcommand{\rzcolor}{$r - z$}

\newcommand{\tractor}{{\tt Tractor}}
\newcommand{\legacypipe}{{\tt Legacypipe}}
\newcommand{\obiwan}{{\tt Obiwan}}

\newcommand{\sextractor}{{\tt Source Extractor}~}

\newcommand{\dev}{de Vaucouleurs}
\newcommand{\healpix}{{\tt HEALPIX}}

\graphicspath{{./}{./figs/}}

\usepackage[dvipsnames]{xcolor}

\title[Removing Imaging Systematics with \obiwan]{Removing Imaging Systematics from Galaxy Clustering Measurements with \obiwan : Application to the SDSS-IV extended Baryon Oscillation Spectroscopic Survey Emission Line Galaxy Sample }

\author[H. Kong et al.]{
Hui Kong,$^{1}$\thanks{E-mail: kong.291@osu.edu}
Kaylan J. Burleigh,$^{2,3}$
Ashley Ross,$^{1}$
John Moustakas,$^{4}$
Chia-Hsun Chuang,$^{5}$
\newauthor
Johan Comparat,$^{6}$
Arnaud de Mattia,$^{7}$
H{\'e}lion du Mas des Bourboux,$^{8}$
Klaus Honscheid,$^{1}$
\newauthor
Sichen Lin,$^{9}$
Anand Raichoor,$^{7,10}$
Graziano Rossi,$^{11}$
Cheng Zhao$^{7}$
\\
$^{1}$Department of Physics, The Ohio State University, 191 West Woodruff Avenue,Columbus, Ohio 43210, USA\\
$^{2}$Department of Astronomy, University of California at Berkeley, 501 Campbell Hall \#3411, Berkeley, CA 94720, USA\\
$^{3}$Lawrence Berkeley National Laboratory, One Cyclotron Road, Berkeley, CA 94720, USA\\
$^{4}$Department of Physics \& Astronomy, Siena College, 515 Loudon Road, Loudonville, NY, USA 12211\\
$^{5}$Kavli Institute for Particle Astrophysics and Cosmology, Stanford University, 452 Lomita Mall, Stanford, CA 94305, USA\\
$^{6}$Max-Planck-Institut f\"{u}r extraterrestrische Physik (MPE), Giessenbachstrasse 1, D-85748, Garching bei M\"unchen, Germany\\
$^{7}$CEA, Centre de Saclay, IRFU/SPP, F-91191 Gif-sur-Yvette, France\\
$^{8}$Department of Physics and Astronomy, University of Utah, 115 S 1400 E, Salt Lake City, UT 84112, USA\\
$^{9}$Center for Cosmology and Particle Physics, Department of Physics, New York University, 726 Broadway,
New York, NY 10003, USA\\
$^{10}$Institute of Physics, Laboratory of Astrophysics, Ecole Polytechnique F\'{e}d\'{e}rale de Lausanne (EPFL), Observatoire de Sauverny, CH-1290 Versoix, Switzerland\\
$^{11}$Department of Astronomy and Space Science, Sejong University, 209, Neungdong-ro, Gwangjin-gu, Seoul, South Korea
}

\date{Accepted XXX. Received YYY; in original form ZZZ}

\pubyear{2020}

\begin{document}
\label{firstpage}
\pagerange{\pageref{firstpage}--\pageref{lastpage}}
\maketitle

\begin{abstract}
This work presents the application of a new tool, \obiwan , which uses image simulations to determine the selection function of a galaxy redshift survey and calculate 3-dimensional (3D) clustering statistics. This is a forward model of the process by which images of the night sky are transformed into a 3D large--scale structure catalog. The photometric  pipeline automatically detects and models galaxies and then generates a catalog of such galaxies with detailed information for each one of them, including their location, redshift and so on. Systematic biases in the imaging data are therefore imparted into the catalogs and must be accounted for in any scientific analysis of their information content.
 \obiwan\ simulates this process for samples selected from the Legacy Surveys imaging data. This imaging data will be used to select target samples for the next-generation Dark Energy Spectroscopic Instrument (DESI) experiment. Here, we apply \obiwan\ to a portion of the SDSS-IV extend Baryon Oscillation Spectroscopic Survey Emission Line Galaxies (ELG) sample. Systematic biases in the data are clearly identified and removed. We compare the 3D clustering results to those obtained by the map--based approach applied to the full eBOSS sample. We find the results are consistent, thereby validating the eBOSS ELG catalogs, presented in \cite{raichoor20a}, used to obtain cosmological results.
\end{abstract}


\section{Introduction} \label{sec:intro}

Galaxy surveys allow astronomers to measure how galaxies cluster at different times in the past. These clustering statistics provide a measure of the expansion rate of the Universe and can answer many other fundamental questions about the Universe \citep{peebles1980}. Some of the most widely known galaxy--redshift surveys include the CfA Redshift Survey \citep{cfaOne, cfaTwo}, The Sloan Digital Sky Survey (SDSS) I and II \citep{sdssYork}, The 2dF Galaxy Redshift Survey \citep{2dFGRS}, WiggleZ \citep{wigglezSurvey}, The SDSS-III Baryon Oscillation Spectroscopic Survey (BOSS) \citep{bossSurvey}, and The SDSS-IV extended BOSS (eBOSS) \citep{dawson2016sdss}. Images of the night sky are transformed into a 2--dimensional large--scale structure (LSS) catalog by passing them through a pipeline that automatically detects and models galaxies and stars in the calibrated images. This becomes a 3--dimensional catalog by selecting galaxies that satisfy particular selection criteria and obtaining spectra and measuring redshifts for them. Redshifts are determined from spectra measured for a subsample of these galaxies. The resulting 3D catalog of galaxy angular positions and redshifts is then used to compute clustering statistics suitable for the extraction of cosmological parameters.

Removing biases and systematics due to the imaging data (imaging systematics) is critical for measuring unbiased clustering statistics. Map--based methods, such as template subtraction and mode projection \citep{biasInTemplateMethod}, have successfully removed imaging systematics from the SDSS, WiggleZ, BOSS, and eBOSS surveys; however, it is unlikely that these methods, in their current state, will be accurate enough for future and on going galaxy surveys, such as the Legacy Surveys. Map--based methods use a pixelization scheme, such as \healpix\, \citep{healpix}, to subdivide the sky into equal--area pixels and then compute various per--pixel quantities. The number of galaxies in the LSS catalog (data) occupying these pixels is compared to, e.g., the average seeing, sky brightness, exposure time, etc. (imaging meta--data) and Galactic foregrounds (e.g., the amount of dust extinction) in each pixel. Correlations between the data and non--data maps are assumed to be due to imaging systematics and are turned into pixel weight maps (in configuration space) or mode weights (in Fourier space). These weights are used to model variations of the angular selection function with imaging properties in the LSS catalogs \citep{biasInTemplateMethod}.

The map-based methods can essentially be divided into ``template subtraction" \citep{sdss1Systematics, myers06, sdss8Systematics, sdss8Companion, sdss9Systematics, sdss12Systematics, wigglezSelectionFunc, delubacSystematics, qsoDepthExtinction, prakashRegressionTech, myersRegressionTech, elvinpoole} and ``mode projection" \citep{rybicki92, tegmark98, uros04, biasInTemplateMethod, leistedt13}. Template subtraction is based on a model for variations of galaxy densities with imaging systematics. Pixel weights are used to correct for the galaxy density fluctuations.  Weights can be applied to randoms or their inverse to data. To avoid modeling chance correlations, only the systematic maps with the largest data cross correlation are included. Mode projection treats the systematic maps as adding noise to each mode in Fourier space or pixels in configuration space, so that values in the data covariance matrix are increased for modes where each systematic map is large. It robustly mitigates the impact of the linear combination of the systematics, but does not include non-linear effects.

The Dark Energy Spectroscopic Instrument (DESI; \citealt{desiScience, desiInstrument}) recently saw first light. It will collect an order of magnitude more galaxy redshifts than are currently publicly available. New methods of data analysis are likely required in order to optimally extract the information while keeping systematic uncertainties sub-dominant. The map-based methods mentioned above are limited by the fact that even though there are dozens of maps that can be created, there is no guarantee the relevant quantities have indeed been mapped. Only the systematic effects known a priori can be modeled. Further, it is standard practice to mask regions, e.g., near bright stars where the imaging data has been corrupted. These masks are binary, whereas the effect on the imaging is unlikely to be a step function in terms of the data quality. The treatment of such issues does not fit neatly into the map-based approach and fully removing affected imaging data would remove untenably large areas from the survey data. For DESI, there is a further complication. It uses imaging data from the Legacy Surveys \citep{overviewPaper}, which is a joint analysis of images from three telescopes. Each telescope obtains multi-- and same--band images of the same part of the sky that are separated by month to year time baselines.

We present a new method for removing imaging systematics at the individual exposure level from future and ongoing surveys that does not require maps of imaging systematics, foregrounds, or other a priori knowledge (i.e. it is non--parametric), and that corrects for biases and systematics in the software pipeline that produced the LSS catalog. We apply our method to DECam data from the Legacy Surveys using the \obiwan\, code \citep{obiwanMethods}. We inject realistic emission line galaxies (ELGs) into the DECam images used to create DR3--era \tractor\, catalogs, from which ELG targets were selected by eBOSS \citep{dawson2016sdss} for spectroscopic follow-up \citep{anand17}. These photometric observations were released as part of SDSS data-release 16 \citep{ahumada2019sixteenth}.

Here, we use \obiwan\, to perform Monte Carlo simulations of how the \legacypipe/\tractor\, pipeline \citep{tractorPaper} detects and forward--models eBOSS ELG--like galaxies and we use the results to support those cosmological analyses. \obiwan\, injects sources into coadded DECam images and builds a LSS catalog using \sextractor. Although the technique of injecting model sources into
imaging and recovering their photometry in the presence of noise has been around for decades (e.g., \citealt{1987PASP...99..191S}), only recently has this method been used to account for
the effect of imaging systematics on measurements of the correlation
function. For example, BALROG \citep{balrog} is used in the Dark Energy Survey (DES \citealt{abbott2019dark}) to correct for imaging systematics. However, \obiwan\, is unique in that it operates on individual exposures and (by virtue of \legacypipe\, and \tractor) maximizes the likelihood of the data to find the best model parameters for each detected source. Benefits of using individual exposures and maximum likelihood (not heuristic) techniques are discussed in \cite{obiwanMethods}.

 Our goal is to apply \obiwan\ over one `chunk' of the eBOSS ELG data and use the results to compute the 3D clustering statistics, comparing the results to those with no correction and to the template-subtracted eBOSS results.
 This is also a preparatory step towards future analyses of the DESI galaxy samples, as DESI will select targets using Legacy Surveys data and its five--year survey is significantly more complicated than eBOSS \citep{desiScience, desiInstrument}.

This study helps support the coordinated release of the final eBOSS measurements of BAO and RSD in the clustering of emission line galaxies \citep[ELG ($0.6<z<1.1$);][]{raichoor20a,tamone20a,demattia20a}. These studies are supported by the mock catalogs described in \citep{lin20a,zhao20a} and the analysis of $N$-body simulations in \citep{Alam20,Avila20}. Analogous efforts for the eBOSS Luminous Red Galaxies (LRG) are presented in \citep{LRG_corr} and \citep{gil-marin20a}, with N-body simulations described in \citep{rossi20a} ; and for eBOSS quasars (QSO) in \citep{neveux20a}, \citep{hou20a}, and \citep{smith20}. The cosmological interpretation of these results, in combination with the eBOSS luminous red galaxy and quasar samples, past SDSS galaxy samples, and in combination with other probes is found in \citet{eBOSS_Cosmology}

This paper is structured as follows. In Section \ref{sec:data}, we describe the imaging and spectroscopic data we use and the eBOSS ELG target selection criteria. In Section \ref{sec:methods}, we summarize how \obiwan\, and \tractor\, work and the algorithms we use for image processing and removing imaging systematics. In Section \ref{sec:Analysis}, we describe the method we use to analyzing our output data, including the 1-point statistics and the correlation function. In Section\ref{sec:results}, we present our \obiwan\, Monte Carlo simulations of the imaging data used to select eBOSS ELGs, and the resulting density map, systematic maps and correlation functions. We conclude in Section \ref{sec:conclusions}. The Appendix presents biases and systematics in the Legacy Surveys image reduction pipeline, and the additional information needed to reproduce our \obiwan\, Monte Carlo simulations.

\section{Data}
\label{sec:data}

\subsection{The DECam Legacy Survey }

The DECam Legacy Survey (DECaLS) is one component of The DESI Legacy Imaging Surveys\footnote{\url{http://legacysurvey.org/}}, which amasses and processes imaging data over 14,000 deg$^2$. DECaLS is 
a \gb, \rband, \zb-band survey of 9,000 deg$^2$ of the southern sky using the Blanco 4-m telescope and DECam camera \citep{DECam} in Cerro Tololo, Chile. DECam has a field of view of 3.18 deg$^2$ and is a mosaic of 62 CCDs, each having 4096x2046 pixels, with pixel scale of 0.262\arcsec\, pixel$^{-1}$. The DECaLS depth is 1--2 mag deeper than the SDSS. For more details see \cite{overviewPaper, burleigh2020observing}. The DR3 data release also includes some Non-DECALs survey data which is also observed by the DECam camera.

\begin{figure}
\begin{center}
       \includegraphics[width=0.98\columnwidth]{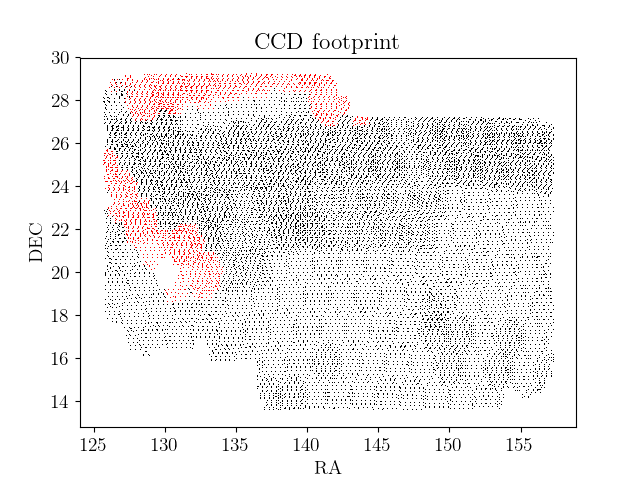}
\caption{The eBOSS NGC `chunk 23' footprint (coordinates are J2000), as traced by the CCDs used for the imaging data. The black points were processed to generate the DECaLS DR3 catalog and red ones were processed by the eBOSS team.}
\label{fig:ccds-eboss}
\end{center}
\end{figure}

The eBOSS ELG target selection \citep{anand17} applied in the region we study used a combination of DR3\footnote{\url{http://legacysurvey.org/dr3}} \tractor\, catalogs and also a set of reprocessed DR3 \tractor\, catalogs, processed by the eBOSS team. This extra data included DECam images observed after the DR3 March 2016 cutoff, and, in order to include this cutoff, which required some re-processing of images included in DR3. We will refer to these reprocessed data as the DR3--plus catalogs. The list of DECam CCDs used to create the DR3--plus catalogs is available online\footnote{\url{/global/cscratch1/sd/huikong/obiwan_Aug/repos_for_docker/obiwan_data/legacysurveydir_dr3/survey-ccds-ebossDR3.fits.gz}}. Fig. \ref{fig:ccds-eboss} shows the locations of the CCDs we study, with the CCDs processed for DR3 \tractor\ catalogs shown in black and CCDs processed by the eBOSS team shown in red.

The model profiles produced from this era of DECaLS data are point source (PSF), an exponential fixed at $r_e=0.45^{\prime\prime}$ (SIMP), exponential (EXP), de Vaucouleurs (DEV), and a composited of EXP and DEV (COMP)\footnote{\url{http://legacysurvey.org/dr3/description/}}. \tractor\ fits sources to each of these models and favors simpler profiles \citep{tractorPaper}.

\subsection{eBOSS} 
\label{sec:eboss-ts}
 SDSS-IV \cite{sdss4} conducts multiple observing programs using the 2.5-meter Sloan Telescope \citep{Gunn06} at the Apache Point Observatory in New Mexico, USA. As part of this program, over a five year period eBOSS collected spectra of quasars, luminous red galaxies, and ELGs in order to make 3D maps of the structure of the Universe.

The eBOSS observations obtained 1000 spectra per observation from fibers plugged into holes in pre-drilled aluminum plates. Spectra were obtained from each fiber using the BOSS double-armed spectrographs \citep{Smee13}, covering the wavelength range 3600 to 10000 \r{A} with R = 1500 to 2600. The eBOSS observations were divided into non-overlapping regions denoted as `chunks'. Within each chunk, the placing of plates was optimized in order to assign the most fibers on targets (with constraints related to, e.g., the minimum fiber separation within a plate being 62$^{\prime\prime}$). ELGs were observed eBOSS chunks 21, 22, 23 and 25. See \cite{dawson2016sdss,raichoor20a,ross20a} for more details. We study the ELG clustering in chunk 23, but use redshift information from chunks 21 and 22.

\subsubsection{ELG Target Selection}

eBOSS selected ELGs from DR3--plus \tractor\, catalogs having clean DECaLS photometry, locations outside bright star masks, sufficient \gb--flux to be [O II] emitters and star forming galaxies, and \grcolor\, and \rzcolor\, color associated with galaxies in the desired redshift range of 0.6 -- 1.1. The regions include 620 deg$^2$ in the South Galactic Cap (SGC) and 600 deg$^2$ in the North Galactic Cap (NGC). ELGs in the SGC are selected using the following \tractor\, catalog cuts, intended to

\noindent only keep sources touching ccds,
\begin{eqnarray}
\label{eqs:elgsel}
 \verb|brick_primary == True|
\end{eqnarray}
 
 \noindent select clean photometry,
 
 \begin{eqnarray} 
 \label{eqs:mask}
 \verb|decam_anymask[grz] == 0|
 \end{eqnarray}

\noindent select [OII] emitters,
 \begin{eqnarray}
\label{eqs:otwo}
 21.825 < g < 22.825
\end{eqnarray}

\noindent and isolate sources to the desired redshift range,
 \begin{eqnarray}
\label{eqs:redshift1}
-0.068\, (r-z) + 0.457 < g-r < 0.112\, (r-z) + 0.773\\
\label{eqs:redshift2}
 0.218\, (g-r) + 0.571 < r-z < -0.555\, (g-r) + 1.901\label{eqs:elgself}
\end{eqnarray}

\noindent The NGC cuts are identical except for,

\begin{eqnarray}
\label{eqs:redshift3}
21.825 < g < 22.9\label{eqs:gcut}\\
\label{eqs:redshift4}
0.637\, (g-r) + 0.399 < r-z\label{eqs:low_z}
\end{eqnarray}

\noindent Bright star masks are also applied. The NGC data includes eBOSS chunks 23 and 25, while the SGC data includes chunks 21 and 22. For more details see \cite{anand17}.

\subsubsection{Our selection of eBOSS data}
We use eBOSS ELG data from the SGC in order to build a model that we apply to the study of ELGs in eBOSS chunk 23. (We do not use data from eBOSS chunk 25.) All eBOSS data used in this study is taken from the catalogs produced by the eBOSS team described in \cite{raichoor20a}.

For selecting data and redshifts to build the \obiwan\ model for the eBOSS population, we use data from the `full' catalogs in the SGC. We select objects with good redshifts and apply some additional cuts on photometric properties. The full selection is described as

\begin{itemize}
\item !NGC
\item \verb|z_ok == 1| 
\item $0 \le \text{redshift} \le 2$
\item $\text{type} \neq \text{COMP}$
\item $\rhalf  < 5.0$\arcsec
\end{itemize}

$r_{half}$ is the half light radius of source profiles.
We will be using the SGC data in order to produce truth catalogs. Only 0.5\% of the data has $\rhalf  > 5.0$\arcsec and we expect these objects to be the result of noise, rather than reflective of the true profile of the population. We drop COMP sources because they comprise less than 1\% of the sample and present modeling complications. This selection provides 113,386 eBOSS ELG redshifts, with an area of 377 $deg^2$. We describe how these data are combined with data from other surveys that have redshifts beyond the selection described above in order to simulate a representative sample of chunk 23 ELGs in Section 

We study the performance of \obiwan\ using data from eBOSS chunk 23. For clustering measurements, we use the `clustering' catalog described in \cite{raichoor20a}. \obiwan\, allows us to estimate the angular selection function in a different way than applied to the standard eBOSS catalogs, which use template subtraction. In order to incorporate the \obiwan\, selection function, an alternative catalog is produced for chunk 23. We describe this further in Section \ref{sec:methods-cf}.

\subsection{Data from outside eBOSS}
\label{sec:elg-like}
A significant component of fluctuations in the eBOSS ELG target density is from varying amounts of targets whose true photometry lies outside of the eBOSS selection cuts but scatters in due to photometric noise. In order to more fully understand the typical redshifts (and thus likelihood of being included in the clustering analysis) of such data, we require spectroscopic samples with selections from outside of the eBOSS criteria. First, we extend the DECaLS selection to include all sources in the SGC region within 0.2 magnitudes of the color and magnitude selections defined by eqns. \ref{eqs:elgsel}-\ref{eqs:elgself}. We denote this as the eBOSS ELG-like sample. We select the number of 0.2 magnitudes as a trade-off between our desire to efficiently simulate the eBOSS ELG sample (which motivates not including objects that have a low chance of scattering into the selection) and our desire to be complete (which motivates including everything with a non-zero chance of selection).

In order to assign redshifts to the data outside of the eBOSS selection bounds, we use data from the DEEP2 \citep{deep2} and VVDS (\citealt{2005A&A...439..845L} \citealt{fevre:hal-01113687}) galaxy redshift surveys, as each has overlap with DECaLS DR3 data in the SGC. DEEP2 and VVDS surveys are described described in more detail below.

\subsubsection{The DEEP2 Galaxy Redshift Survey (DEEP2)}
\label{sec:deep2}
DEEP2 obtained about 50,000 high resolution (R $\sim 6000$) spectra of redshift $\sim 1$ galaxies using the DEIMOS multi--object spectrograph on Keck 2. The DEEP2 footprint is 2.8 deg$^2$, split into four disjoint regions: Field 1 (14hr), Field 2 (16h), Field 3 (23h), and Field 4 (02h). We create a DEEP2 (DR4) and DECaLS DR3 matched table by finding the nearest DR3 \tractor\, catalog source within a 1\arcsec\, search radius of each DEEP2 spectrum. The DECaLS DR3 footprint does not overlap Field 1, so our table only includes Fields 2--4. We refer to it as the DR3--DEEP2 table and use it in Section \ref{sec:injecting-elgs}. We obtain redshifts for 1065 EXP and 86 DEV galaxies after matching to the eBOSS ELG-like catalog. 

\subsubsection{The VIMOS-VLT Deep Survey (VVDS)}
\label{sec:vvds}
VVDS contains three complementary surveys: The VVDS-Wide, the VVDS-Deep and the VVDS-Ultra-Deep. We use one field in the VVDS-Wide survey, the VVDS-22h field. It has a total area of 4.0 $deg^2$, and a total of 13291 identified objects around redshift of 0$\sim$2. We create a VVDS and DR3 matched table using the same procedure as for the DEEP2 survey, see section \ref{sec:deep2}. We refer to this as the DR3--VVDS table. We also combine DR3--DEEP2 table with DR3--VVDS table to form a combined table, and we call it the DR3--DEEP2--VVDS table. We obtain redshifts for 786 EXP and 65 DEV galaxies after matching to the eBOSS ELG-like catalog. 

\subsubsection{DR3-DEEP2-VVDS sample}
For the ELG-like sample we find that DEEP2 and VVDS provide redshifts for about 30\% of the DECaLS sources within the overlapped footprint. The distribution of colors and magnitudes for the DEEP2 and VVDS samples fully overlap the ELG-like color/magnitude space and thus we treat any incompleteness via re-sampling as described in subsequent sections.
We call our combined result the DR3--DEEP2--VVDS sample. The final DR3--DEEP2--VVDS sample has obtained redshifts for 1,851 EXP and 151 DEV galaxies. With the combination, we get an adequate volume of samples outside the eBOSS ELG selection box with reliable redshifts. For each of these sources, we match to the nearest eBOSS ELG-like source using a 1\arcsec\ matching radius.
We then apply the following cuts, 
\begin{eqnarray}
0 \le \text{redshift} \le 2 \\
\rhalf  < 5.0\arcsec.\\
\end{eqnarray}

The DR3--DEEP2--VVDS galaxy sample fulfills the purpose of providing redshifts to galaxies from outside the eBOSS ELG selection cuts. This will allow us to later characterize the type of objects we expect to scatter (e.g. from \tractor\, measurement error) across the ELG selection boundaries and into the eBOSS ELG sample.

\section{\obiwan}
\label{sec:methods}

\subsection{Basic Methodology}
\label{sec:methods-obiwan}

\obiwan\, is fully described in \cite{obiwanMethods}. We repeat the essential details here. We add simulated sources (inject), with properties closely matched to the galaxies of interest, to random locations in the imaging data within the relevant survey mask. We uniformly inject sources over the footprint of a given survey and measure their photometry using the same pipeline as for the real data. The fluctuations in the number of sources that pass the same selection as the given survey
thus trace its angular selection (with some Poisson noise related to the density at which we sample a given area). This selection function can be applied when calculating clustering statistics.

\obiwan\, modifies the \gb, \rband, \zb\, images that \legacypipe\, operates on by adding simulated sources to the individual exposures and appropriately modifying the inverse variance images. These simulated sources are the same as model sources that \tractor\, creates when fitting real sources. This means that in the case of no noise added, the simulated sources will be perfectly fit by \tractor, and thus any output parameter of the fitted simulated source will match the input parameter. Our tests show that the changes between parameters are subtle compared to the total changes in a simulation run on a real image. The simulated sources include Poisson noise from the source itself. The power of \obiwan\, is that the injected sources inherit the sky background, systematics, or whatever else is present in the data, so nothing more than the simulated galaxy or star of interest is injected. \legacypipe\, does not know the images have been modified and source detection, model fitting, and model selection proceed as usual. 

\obiwan\, performs a Monte Carlo Simulation by injecting the simulated galaxies at random right ascension (RA) and declination (Dec), running \legacypipe, and repeating for the same images as what are used to produce the eBOSS ELG imaging data. Blending can occur between pairs of real--real, real--simulated and simulated--simulated sources. Our goal is to simulate effects involving galaxies, so we prevent blending between simulated--simulated sources. We temporarily set aside all simulated sources that would be within 5\arcsec\, of another simulated source, and injected those set--aside sources during the next Monte Carlo iteration. Blending between real and simulated sources is allowed (and needed to fully simulate the angular selection function). This 5\arcsec\, criterion only applies to pairs of simulated sources. The initially random fluctuations in source density are modified by the geometry of the footprint, source detection, measurement, target selection, and any biases and systematics in the \legacypipe\, pipeline. We will refer to these as \obiwan--randoms, and the truly random galaxy positions (e.g. the RA, Dec for all the sources we inject into the imaging data) as uniform--randoms.

\begin{figure*}
\begin{center}
 
 \includegraphics[width=1.5\columnwidth]{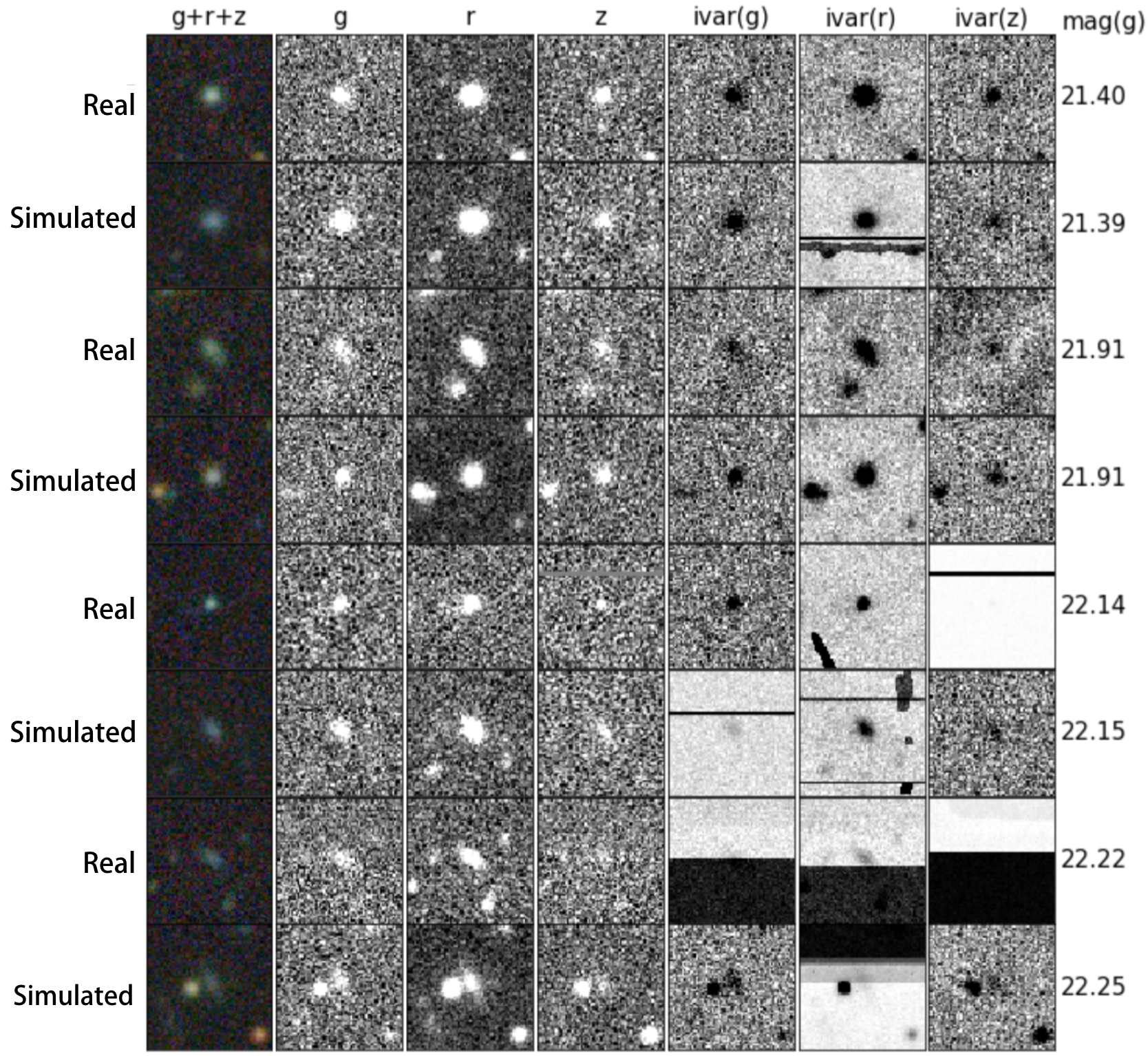}
 \caption{A comparison between real and simulated galaxies that have exponential profiles and relatively bright \gb--band magnitudes. The label for each image is on the left and its corresponding \gb\, magnitude is the number on the right. Each row is a single galaxy. The first column is a three color jpeg for easy visualization. The remaining columns are the per--band full resolution coadds for the \gb, \rband, \zb\ images and associated inverse variance maps. Consecutive rows of Real and Simulated (rows 1 and 2, 3 and 4, etc.) have similar \gb\, magnitude for a fair comparison. Some color difference between corresponding panels is noticeable because they are matched by $g$ magnitude only (not by color).}
 \label{fig:examples-bright}
 \end{center}
\end{figure*}

Fig. \ref{fig:examples-bright} compares real and simulated galaxies that have exponential profiles and relatively bright \gb--band magnitudes. 
Their color and high signal-to-noise (S/N) are not representative of the full distribution. In this sample, we cannot tell the difference between the real and simulated ELGs in the $g$ band image, which is the band that we intentionally set to be very similar. We naturally expect to have a more difficult time telling the difference between renderings at lower signal to noise.

\subsection{Injecting Realistic eBOSS ELGs}
\label{sec:injecting-elgs}

This section summarizes how we generate the representative sample of eBOSS ELG-like galaxies that we inject into the images. The representativeness of our sample is crucial for our method to be statistically unbiased. The representative sample is a distribution of photometric properties, shapes and redshifts. The photometric properties and shapes are sampled from redshifts in order to determine the model parameters defining the random potential ELG targets that are injected into the images.

\begin{figure*}
\begin{center}
 \includegraphics[width=1.9\columnwidth]{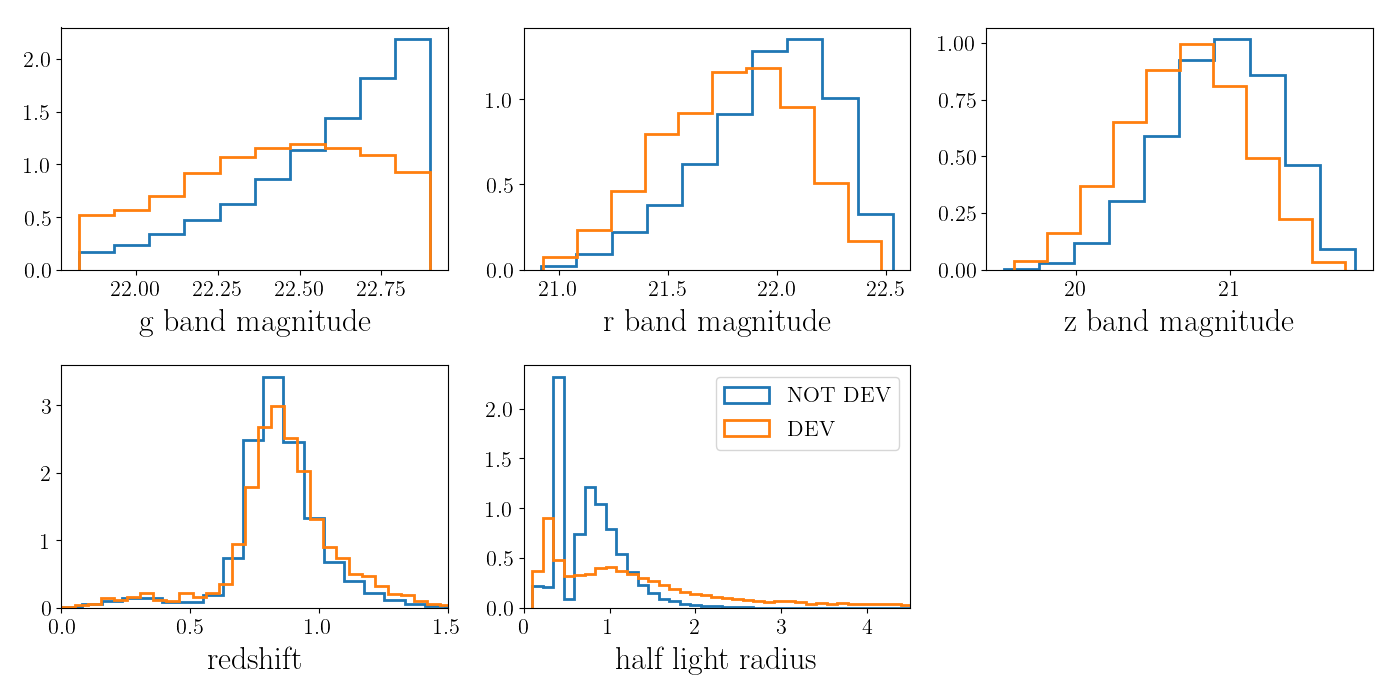}
 \caption{ The PDFs of the normalized distributions \gb, \rband, and \zb\, magnitudes (top row), the redshifts (bottom left), and $\rhalf$ (bottom right) for eBOSS ELGs. The PDFs are split between sources that {\textsc legacypipe} applies de-Vaucouleurs (DEV; orange) and Exponential or PSF (NOT DEV; blue) profiles to in order to determine the photometry. Galaxies fit with DEV profiles are systematically larger and brighter than those fit with EXP/PSF profiles.}
 \label{fig:dev-brighter}
 \end{center}
\end{figure*}

We start the process by studying the properties of the eBOSS SGC data. Morphologically, there are 5\% PSF, 25\% SIMP, 58\% EXP, 11\% DEV, and 1\% COMP galaxies. The majority of the data can be described as having an EXP profile. We assume that all sources that \tractor \, classifies as type PSF are compact and/or unresolved galaxies. These sources have a pixelized PSF profile, which is mathematically equivalent to an exponential  profile having $r_{half}=0$, so we reclassify them as such. We also reclassify SIMP sources as EXP because SIMP are EXP galaxies with fixed rhalf of 0.45. After such consideration, we have 102,100 EXP and 11,286 DEV eBOSS galaxies with eBOSS redshifts. Comparing their properties in terms of redshift, shape, and \gb, \rband, \zb\, flux, we find that DEV galaxies are systematically larger and about 1 mag brighter than EXP in all bands (see Fig. \ref{fig:dev-brighter}). Given these systematic differences, we model the two populations separately and inject them into the true images such that 90\% of the injections are EXP.

\begin{figure*}
\begin{center}
     \label{fig:dr3dp2-vs-eboss-exp}{%
       \includegraphics[width=0.97\columnwidth]{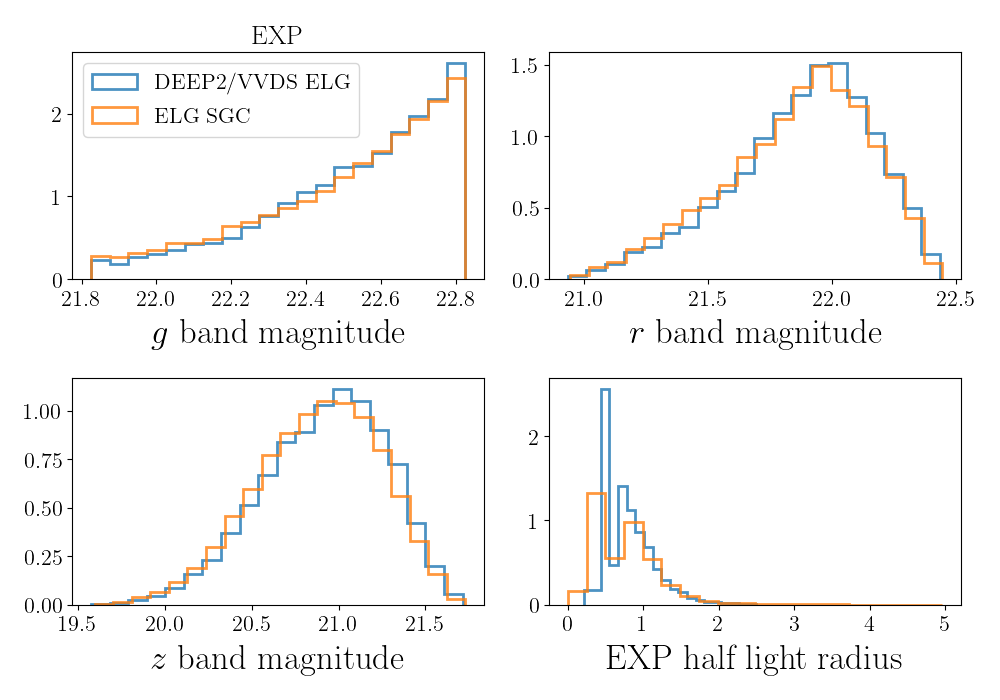}
     }
     \hfill
     \label{fig:dr3dp2-vs-eboss-dev}{%
       \includegraphics[width=0.97\columnwidth]{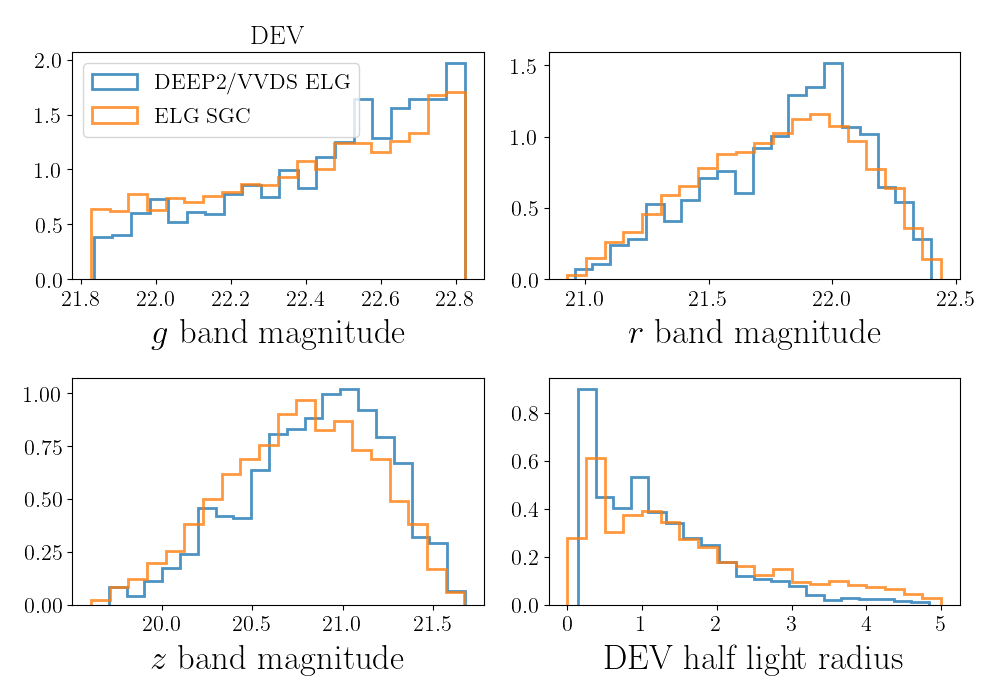}
     }
\end{center}

\caption{Comparison of DECaLS DR3 photometric properties of our eBOSS SGC ELG ("ELG SGC") and our combined sample of DEEP2 and VVDS galaxies selected pass the eBOSS ELG color/magnitude cuts ("DEEP2/VVDS ELG"). We split the samples into those that \legacypipe\, fits with Exponential (EXP; left) de-Vaucouleurs (DEV; right) profiles. The DEEP2/VVDS ELG sample for DEV galaxies is noisy because there are only 151 DEV galaxies in the sample.}

\label{fig:dr3dp2-vs-eboss}
\end{figure*}

We next develop a method to sample from the eBOSS redshift distribution $dN/dz$ using the DR3-DEEP2-VVDS sample. Though this sample distribution does not represent the full population of eBOSS ELG-like galaxies, we weight DR3-DEEP2-VVDS objects such that their redshift distribution matches the one of the eBOSS ELG-like galaxies. Fig. \ref{fig:dr3dp2-vs-eboss} demonstrates that the photometric properties of the redshift-weighted DR3-DEEP2-VVDS sample within the eBOSS ELG box matches that of the full ELG sample. As discussed in Appendix \ref{sec:eboss-almost-targets}, the photometric distribution outside the eBOSS ELG selection box does not have a perfect match. This is because the sampling with redshift down weights bright sources. However, Fig. \ref{fig:dr3dp2-redshift} shows that the output \obiwan-ELG distribution share similar properties as eBOSS ELGs, and this difference does not bias our result.

The redshift-weighted DR3-DEEP2-VVDS sample thus represents a discrete distribution that can be sampled in order to reproduce the properties of eBOSS ELGs. In order to do the sampling, we first fit a 10 component Gaussian mixture model (GMM) to the eBOSS $dN/dz$, clipped to be in the range $0 < z < 2$. This provides an analytic statistical distribution that we can sample from. 

For each random RA,DEC position, we first determine whether the source will be EXP (with 90 per cent probability) or DEV (with 10 per cent probability).
We sample a redshift from the GMM and then we find the nearest redshift for an EXP or DEV source in the DR3--DEEP2--VVDS distribution described above. If the color/magnitudes of the matched DR3--DEEP2--VVDS galaxy lie within the eBOSS target selection, we then find the nearest redshift in the EXP or DEV eBOSS distribution and assign the source the DECaLS model parameters of this same eBOSS ELG. If it is outside the eBOSS target selection cuts, we use the DECaLS model parameters from the original DR3--DEEP2--VVDS source used for this target selection test. This method thus returns a sample that simultaneously matches the distribution of photometric properties in the ELG-like sample and the redshift distribution of the eBOSS sample. After the model parameters are determined, the \obiwan\ source is rendered and injected at the random RA,DEC position.

\subsection{Image processing with \obiwan}

As described in the previous sub-sections, we use \obiwan\, to inject simulated galaxies into the DECam CCDs used to create the DR3-plus \tractor\, catalogs. We use the dr5 version of \legacypipe\,  and not the dr3 version that actually created the DR3-era \tractor\, catalogs. The dr3 version of legacypipe has three sub-versions: dr3a, dr3c, dr3e. eBOSS ELG chunk21,22 are produced by dr3a, dr3c, and eBOSS ELG chunk23 is produced by dr3c, dr3e. The dr5 version is very similar to to the dr3c, dr3e versions of legacypipe, so we perform our simulation on chunk23 eBOSS ELG region. This allows us to apply one version (dr5) of legacypipe to obtain our results.

\section{Analysis}
\label{sec:Analysis}
\subsection{Maps and 1-point statistics}

We use \healpix\ \citep{healpix} to divide the sky into 12$N_{SIDE}^2$ equal area pixels. This is a convenient way to map the density of galaxies and compare it to the \obiwan\ predictions. 

We are also able to map observational parameters that might modulate the target density ($sys$), such as the mean seeing or the total image depth in a given band. The \healpix\, maps allow us to determine the mean densities of galaxies and randoms as a function of these parameters. Defining the ratio of the number of randoms to the number of galaxies $\alpha = N_{ran,tot}/N_{gal,tot}$, we can determine the relative density $\alpha N_{gal}(sys)/N_{ran}(sys)$, where 1 is the expected null result. In practice $sys$ will represent some range of values in the systematic map. We estimate an error on this assuming galaxy and random counts are Poisson distributed
\begin{equation}
\label{equ:sigma_diag}
\sigma = \alpha\left(\frac{n_{gal}(sys)}{n^2_{ran}(sys)}+\frac{n^2_{gal}(sys)}{n^3_{ran}(sys)}\right)^\frac{1}{2}.
\end{equation}

Naively, this ignores the cosmic variance component of the galaxy fluctuations. However, it was demonstrated in \cite{sdss12Systematics} that if one weights galaxies and randoms by $w_{FKP}$ when counting $n_{gal}(sys)$, one recovers approximately the same result as when deriving the uncertainty from many realizations of the data with the same number density and similar clustering properties (i.e., `mock' galaxy samples). We thus apply the $w_{FKP}$ provided in the ELG LSS catalogs \citep{raichoor20a}.

\subsection{The Correlation Function}
\label{sec:methods-cf}

The correlation function is a statistic that measures the clustering of galaxies, relative to a random distribution, for a range of galaxy--galaxy separations \citep{peebles1980, corrfuncHamilton, weinberg, corrfuncErrors, corrfuncIC, corrfuncEboss}. We determine the redshift-space two point correlation function, $\xi$ using the standard \cite{landy93} estimator. In the following analysis we assume a flat $\Lambda$CDM cosmological model with and $\Omega_m = 0.31$. 

\begin{align}
\label{equ:landy93}
\xi(s,\mu) = \frac{DD(s,\mu)-2DR(s,\mu)+RR(s,\mu)}{RR(s,\mu)}
\end{align}

\noindent where $s$ is the separation between two galaxies, and $\mu$ is the the cosine angle between the pair of galaxies and the line-of-sight. We calculate this function numerically in evenly spaced bins of width 5$h^{-1}$Mpc in $s$ and 0.01 in $\mu$. We transform this 2-variable function into one variable by computing the monopole ($\ell=0$), quadrupole ($\ell=2$) and hexadecapole ($\ell=4$), we project this correlation function onto the basis of legendre multipoles $L_{\ell}$ following:

\begin{align}
\label{eqn:poles}
\frac{2\xi_{\ell}(s)}{2\ell+1}=\sum_{i=1}^{100}0.01\xi(s,\mu_i)L_{\ell}(\mu_i)
\end{align}

\noindent In the linear model for redshift-space distortions \citep{Kaiser87}, these three multipoles contain the full clustering information. 

The angular separation ($\theta$) between a pair of points with (RA$_1$, Dec$_1$) and (RA$_2$, Dec$_2$) is,
\begin{equation}
\cos(\theta) = \cos(\psi_1) \cos(\psi_2)\cos(\upvarphi)  + \sin(\psi_1)\sin(\psi_2),
\end{equation}

where $\cos(\upvarphi)$ is defined as:
\begin{equation}
    \cos(\upvarphi) = \cos(\phi_1)\cos(\phi_2)+\sin(\phi_1)\sin(\phi_2) 
\end{equation}
\noindent where $\psi = (-\text{Dec}+90) \,\pi/180$ and $\phi = \text{RA} \times \pi/180$. The s, $\mu$ in correlation function $\xi(s,\mu)$ is defined as 
\begin{align}
s = \sqrt{r_1^2+r_2^2-2r_1{\cdot} r_2{\cdot}cos(\theta)}
\end{align}
\begin{align}
\mu = \frac{|r_1-r_2|}{s}
\end{align}

\noindent $r_1$, $r_2$ here are the distances from the galaxy to the observer.

The eBOSS team created a set of eBOSS ELG randoms for computing clustering statistics of eBOSS ELGs. The full details are provided in \cite{raichoor20a}. We use these on chunk 23 for all $\xi$ calculations. The eBOSS team also provides a set of weights (again described in \citealt{raichoor20a}) to be applied to the galaxies and randoms when determining pair-counts. These include: $w_{\rm sys}$, to correct for imaging systematics; $w_{\rm CP}$, to correct for fiber collisions; $w_{\rm NOZ}$, to correct for redshift failures; and $w_{\rm FKP}$, to more optimally weight the information as a function redshift\footnote{In the catalogs, these columns are: `WEIGHT\_SYSTOT', `WEIGHT\_CP', `WEIGHT\_NOZ', `WEIGHT\_FKP'.}. These weights are simply multiplied by each other and the total weight is applied each galaxy/random when counting pairs. We will use \obiwan\ to produce and test an alternative to $w_{\rm sys}$.

We use EZmocks \citep{10.1093/mnras/stu2301} to compute error bar for our correlation functions. EZmocks are fast mocks that is used to accurately predict the variance of eBOSS ELGs by making multiple realizations of the Universe. We construct our covariance matrix with 1000 EZmocks. The elements for our covariance matrix are defined as 

\begin{align}
\label{eqn:cov_matrix}
Cov_{ij} = \frac{1}{999} \Sigma_{k=1}^{k=1000}(\xi_{ik} - \overline{\xi}_{i})*(\xi_{jk} - \overline{\xi}_{k}),
\end{align}

\noindent where $\xi_{ik}$ is the $i^{th}$ element in the $k^{th}$ EZmock correlation function. $\overline{\xi}_{i}$ is the $i^{th}$ element of the average of 1000 EZmocks. This $\xi$ can be either monopole, quadrupole or hexadecapole. 
The EZmocks created for eBOSS DR16 analysis are described in \cite{zhao20a}.

We will compare the consistency of results using $\chi^2$ values. Given a covariance matrix $C$ for some data vector $D$ and a model data vector $M$
\begin{equation}
    \chi^2 = (D-M)^TC^{-1}(D-M)
\end{equation}

\noindent The covariance matrix we use is either the diagonal matrix described in equation \ref{equ:sigma_diag}, or the full covariance matrix produced by mocks in equation \ref{eqn:cov_matrix}.

\section{Results}
\label{sec:results}

We inject 1.2M simulated galaxies, at a density of 3200 per deg$^2$ into the eBOSS CCDs for the NGC region. Due to the removal of simulated galaxies within 5\arcsec\, of other simulated galaxies, the mean becomes 3152 per deg$^2$. About 20 per cent, or 646 per deg$^2$, of these injected galaxies have true parameters that pass the eBOSS NGC ELG target selection. The eBOSS ELG target density in the NGC is 200 per deg$^2$, so our injected sample (before source detection and \tractor, measurement) has 3.2x the density of the real galaxy sample. 20.4\% of the \obiwan-randoms passe our final ELG selection function, so the total number of ELGs in \obiwan-randoms (after source detection and \tractor, measurement) is very similar to the total number of injected ELGs.

In the subsections that follow, we go through particular aspects of the eBOSS chunk 23 \obiwan\, results. We first compare the color/magnitude distributions of the outputs compared to eBOSS data and to the input truth. We then study how well the \obiwan\ results predict the fluctuations in target density. Finally, we apply the \obiwan\ results to the clustering measurements.

\subsection{Color/Magnitude distributions}

\begin{figure*}
\begin{center}
 \includegraphics[width=1.7\columnwidth]{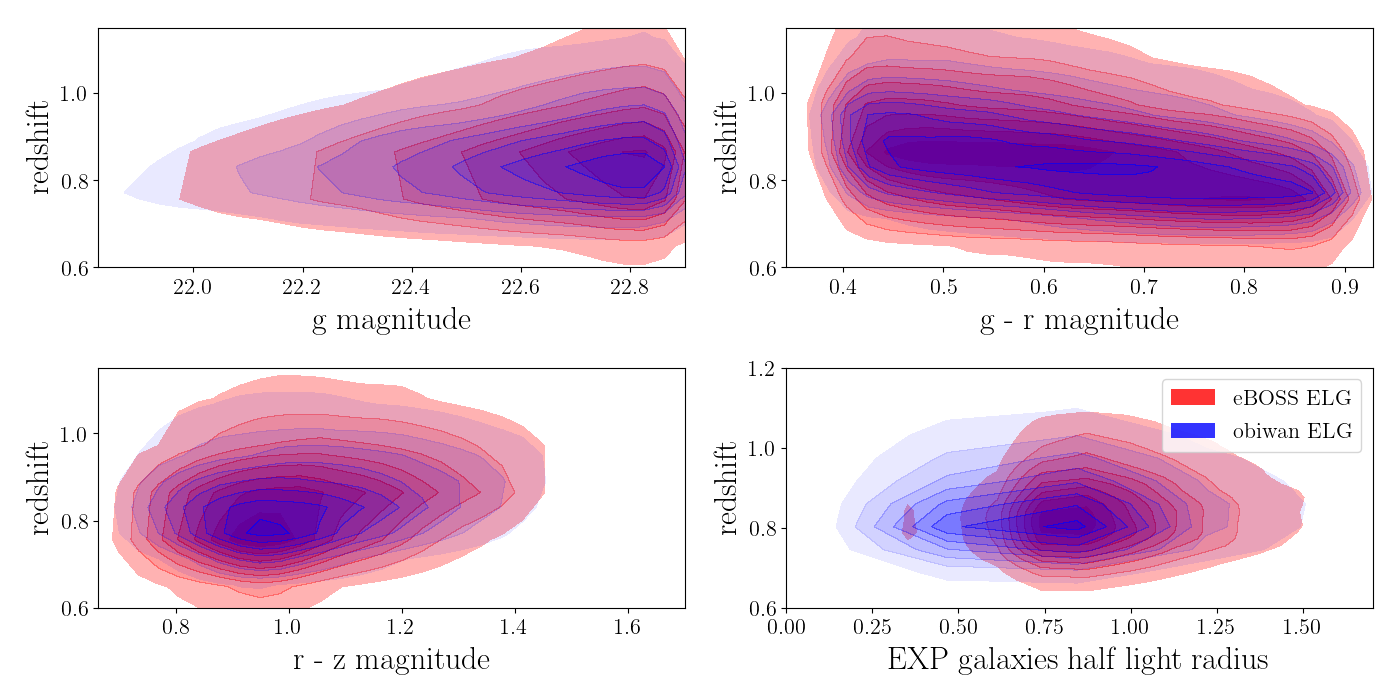}
\end{center}

 \caption{2-dimensional contour plots showing how redshift depends on \gb, \rband, \zb\, mag and $\rhalf$ for galaxies \legacypipe\, fits with an Exponential (EXP) profile in the eBOSS (blue) and Obiwan (simulated eBOSS; red) samples. There is a clear mismatch in obiwan ELG half light radius; as we desribe in the text, this does not have a strong correlation with ELG recovery rate, and will not strongly affect our sample.}
 \label{fig:dr3dp2-redshift}
\end{figure*}

Fig. \ref{fig:dr3dp2-redshift} shows the degree to which \obiwan\ ouputs agree with the eBOSS colour/magnitude/redshift distributions. Here, \obiwan\ ELGs are \obiwan-randoms whose measured photometry is within eBOSS ELG target selection box described in \ref{sec:eboss-ts}. The joint magnitude and redshift distributions of real and simulated targets are similar, implying our simulated targets have similar properties to the real ELG targets. However the distribution of \obiwan\, ELGs $r_{half}$ (half light radius) is quite different from eBOSS ELGs. The reason is that the DR5 version of \legacypipe\, is biased towards greater $r_{half}$ measurements. However, we find this effect is minor compared to other sources of fluctuation in the recovery rate.  Fig. \ref{fig:dev-brighter} shows that \dev\, galaxies are more extended than exponential galaxies. Fortunately, this does not bias our results, as we find \legacypipe\, is equally good at recovering exponential and \dev\, sources. The injected galaxies are 90\% exponential and 10\% \dev\,, and \legacypipe\, recovers 97\% of the exponentials and 96\% of the \dev. Basically, $r_{half}$ is not an influential factor in eBOSS ELG properties, so it has a minor effect in our final result. For more details see Appendix \ref{sec:biases-systematics}. 

\begin{figure*}
\begin{center}
 \includegraphics[width=1.7\columnwidth]{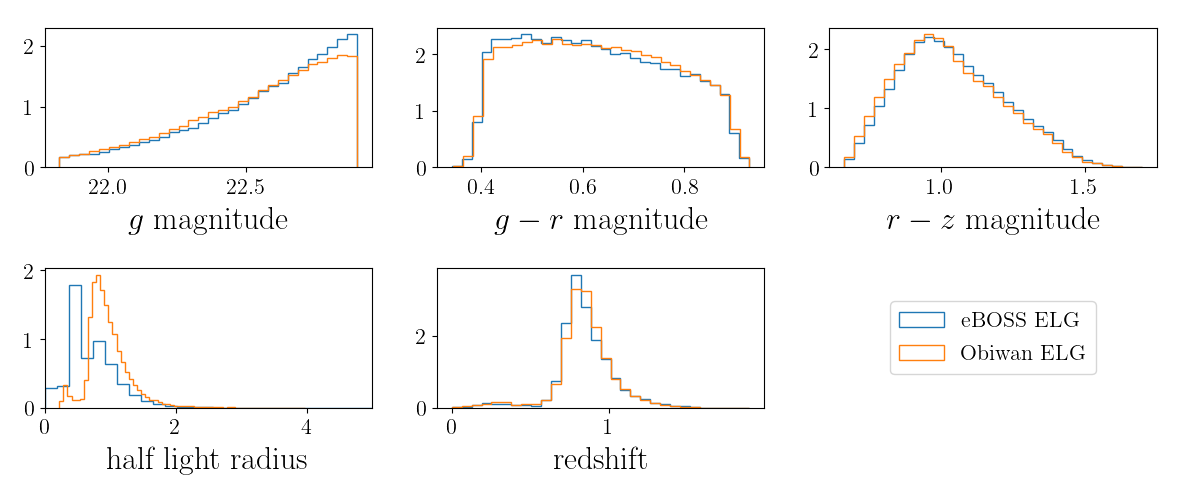}
\end{center}

 \caption{PDFs comparing the photometric properties and redshifts of eBOSS ELGs (blue) and the Obiwan ELGs (orange) meant to simulate the eBOSS sample. 
 }
 \label{fig:input}
\end{figure*}

Fig. \ref{fig:input} shows the \gb, \rband, and \zb\, mag histograms for the Obiwan-ELG galaxies and the real eBOSS ELG galaxies. We can see a slight difference in the photometric distributions, most significantly that of the $g$-band. This could be an effect of a bias in this version of
\tractor's\, flux measurement. This is studied further in the Appendix, where Fig. \ref{fig:num-std-dev-and-dmag} shows a slight correlation between $g$ magnitude and magnitude bias. Objects with faint magnitudes are measured to be even fainter, resulting in a density decrease after applying the color cut. Another reason is that all of the true magnitudes we use are within 0.2 magnitudes of the eBOSS selection boundaries. Thus, it is not possible for objects with $g > 23.025$ to scatter into our selection. At the faint end, we find that 10\% of the input sources have $g$-band magnitudes that change by 0.2 magnitudes, and we are thus missing these objects for which $\Delta$ mag would be > 0.2. Despite this imperfection, we will find that the Obiwan-ELG galaxies reproduce the fluctuations observed in the real eBOSS ELG galaxy sample.

We classify the sources in the catalog in three ways: true positives, false positives, and true negatives. True positives (recovered ELGs) are simulated ELGs that remain eBOSS ELGs using \legacypipe's measurements for them. False positives (contaminants) are simulated non--ELGs that pass target selection after \legacypipe\, measures their magnitudes. True negatives (lost ELGs) are simulated ELGs that are either not detected (non--detections) or have sufficient \tractor\, measurement error to fail target selection (measurement--error). 

\begin{figure*}
\begin{center}
 \includegraphics[width=2\columnwidth]{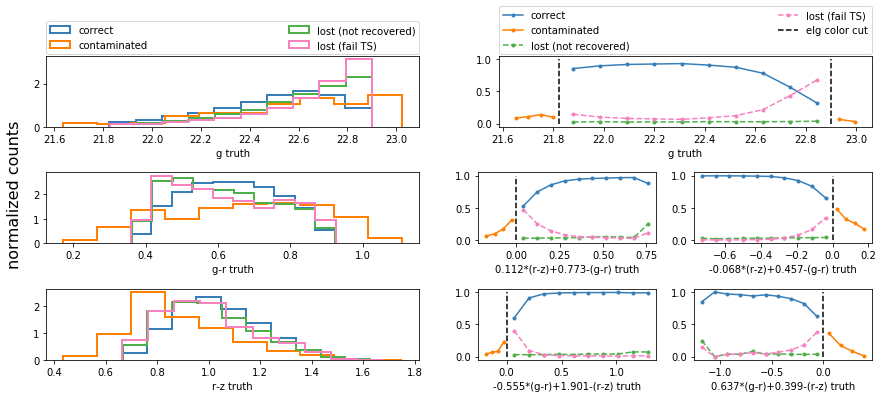}
\end{center}

 \caption{Left: True color and magnitude histograms for the \obiwan\, sources. We display the following possibilities: `correct' denotes that both the true and output \obiwan\, photometry passes the eBOSS ELG cuts; `contaminated' denotes that the output photometry passes the eBOSS ELG cuts but the true photometry does not; `lost (not recovered)' denotes that the source was not detected by \obiwan\,; `lost (fail TS)' denotes that the output \obiwan\, photometry did not pass the eBOSS ELG cuts but the true photometry did. Right: The same categories are shown displaying the ratio of each type of source to the input combined color or magnitude histogram.
We observe that ELGs lost to measurement–error are, on average, the faintest of the simulated galaxies in $g$, $r$, and/or $z$. Contaminants and ELGs lost to non–detections are a minority of the sample and have similar $g$, $r$, $z$ magnitude distributions.  
 }
 \label{fig:rec-lost-contam-mag}
\end{figure*}

Fig. \ref{fig:rec-lost-contam-mag} shows the \gb, \rband, \zb\, magnitude distributions for the recovered ELGs, contaminants, and lost ELGs. Measurement-error is the primary way that ELGs are lost. ELGs lost to measurement-error are, on average, the faintest of the simulated galaxies in \gb, \rband, and/or \zb. This is natural due to the faintest sources having the greatest uncertainty on their measured magnitudes. The issue is further exacerbated by the photometric measurement bias described in Appendix \ref{sec:biases-systematics}. Flux measurements are systematically smaller, and there are a lot of ELGs at high $g$ magnitude range, so a lot of sources get cut off by the $g$ magnitude cut at faint end (Equation \ref{eqs:gcut}). Contaminants and ELGs lost to non-detections are a minority of the sample and have similar \gb, \rband, \zb\, mag distributions.

\begin{figure}
\begin{center}
 \includegraphics[width=\columnwidth]{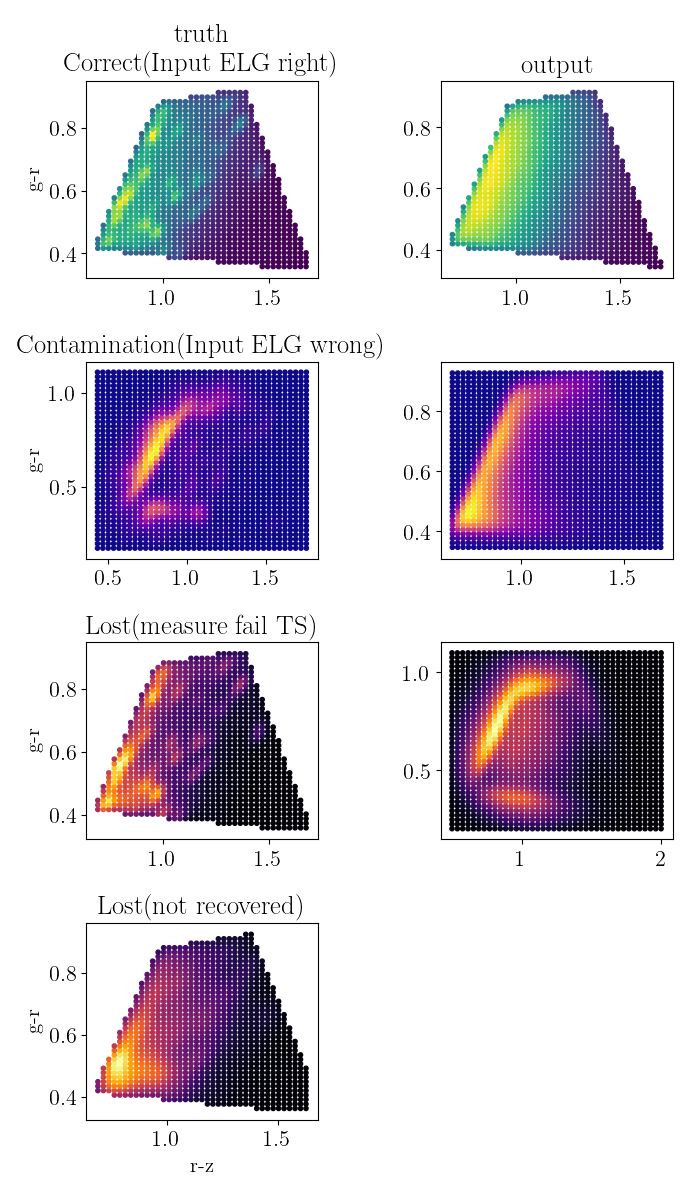}
\end{center}

 \caption{Distributions of \obiwan\, sources split into recovered ELGs, contaminants, and lost ELGs for the eBOSS color box, using the same definitions as Fig. \ref{fig:rec-lost-contam-mag}. The left-hand plot show the true colors of the sources. On the right, we show the output \legacypipe\, measured color.  }
 \label{fig:rec-lost-contam-color}
\end{figure}

Fig. \ref{fig:rec-lost-contam-color} shows the colors for recovered ELGs, contaminants, and lost ELGs. There are some striped patterns in the `truth correct' and `truth lost' panels. This is due to the discreteness of our dr3--DEEP2--VVDS sample. The top right panel shows the eBOSS color box. We can see that most contaminants start at top left of the color box and scatter by $\sim 0.25$ mag to redder \rband-\zb\, colors. This corresponds to a redshift cut (Equation \ref{eqs:low_z}), which means that most contaminants come from low redshift sources. The colors of ELGs lost due to non-detections are distributed over the lower left corner of the ELG selection box. The redshifts in this region are systematically higher. The `tractor contamination' panel shows that there are also a lot of sources scattering into this corner. Combining with the histograms in Fig. \ref{fig:rec-lost-contam-mag}, we know that there are a lot more sources scattering into than outside of this corner. 

\begin{figure*}
 \includegraphics[width=1.7\columnwidth]{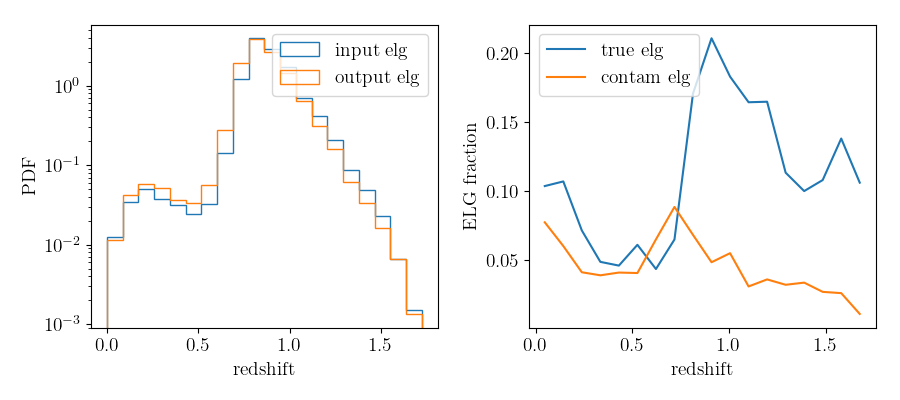}
 \caption{Here we show how the process of measuring photometry with \legacypipe\, modifies the expected redshift distribution. The left-hand panel displays the redshift PDF for injected \obiwan\, galaxies (blue) compared to the same after applying the eBOSS ELG cuts to the \obiwan\, output photometry (orange). The right-hand panel displays the fraction of \obiwan\, sources with outputs passing the eBOSS ELG cuts in each redshift bin that had input `true' photometry passing the eBOSS ELG cuts (blue) and those that did not (`contaminants'; orange). Contaminants primarily enter at redshift range: $0 < z < 0.75$}
 \label{fig:redshifts-recovered}
\end{figure*}

We inject ELGs with the appropriate correlations among brightness, shape, and redshift. Fig. \ref{fig:redshifts-recovered} shows how the injected redshift distribution dN/dz is modified by \legacypipe. The left panel shows that output \obiwan-ELGs are systematically of lower redshift than input obiwan-ELGs. The left panel shows that contaminants primarily enter at the redshift range: $0 < z < 0.75$.

\subsection{Spatial fluctuations}
 
\begin{figure}
\begin{center}
 \includegraphics[width=\columnwidth]{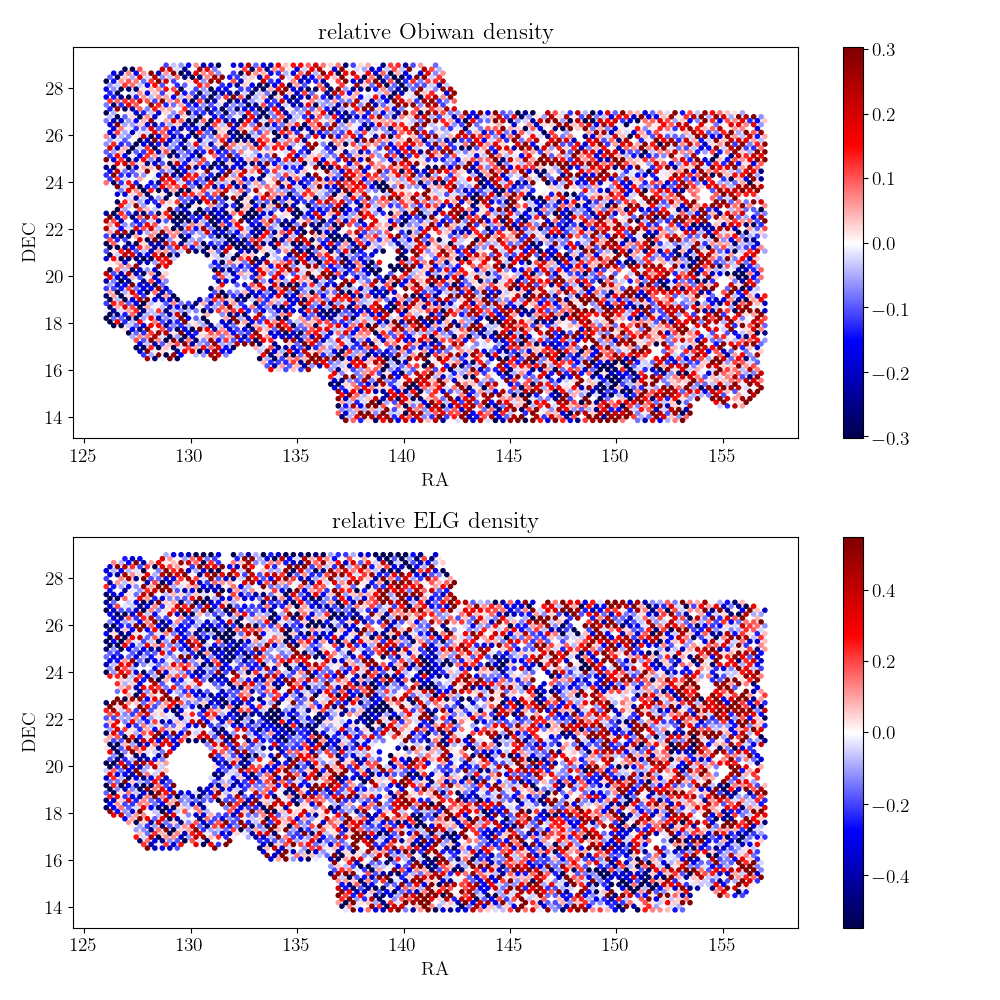}
\end{center}
 \caption{The density distribution of \obiwan\, ELGs (with output photometry passing the eBOSS ELG cuts; upper-panel) and eBOSS ELGs (lower-panel). Looking by eye, we observe similar patterns, e.g., density increasing with RA. This implies that \obiwan\, is successfully predicting large-scale patterns in the recovered ELG density.
 }
 \label{fig:density-plot}
\end{figure}

Fig. \ref{fig:density-plot} displays the number density distribution of \obiwan-ELGs and eBOSS ELGs. One can see that they share similar large-scale patterns. In particular, there is an increase in density with RA. This is a visual demonstration of the power of \obiwan\,. The maps share the same large-scale patterns since they are extracted from the same images, and thus share the same imaging systematics.

In \cite{raichoor20a}, a linear regression is performed testing the eBOSS ELG density against a number of maps of foregrounds and imaging properties that have the potential to systematically bias the number of eBOSS ELGs selected from the photometry. The coefficients of this regression produced the $w_{\rm{sys}}$ to be used for clustering measurements. If \obiwan\, works properly, it should predict the dependency between the eBOSS ELG density and these maps. We 
tested how well \obiwan\, does so by computing ratio of the normalized total number of galaxies $n_{gal}(i)$ to either normalized total number of \obiwan \, or uniform randoms $n_{ran}(i)$, in 10 bins of the value of each map. For a null result, we expect the value in each bin to be consistent with 1. We also weight the ELGs by $w_{\rm{sys}}$ instead. As the $w_{\rm{sys}}$ values were obtained already via a linear regression with these maps, we expect a null result in this case.

\begin{figure*}
\begin{center}
\includegraphics[width=\textwidth]{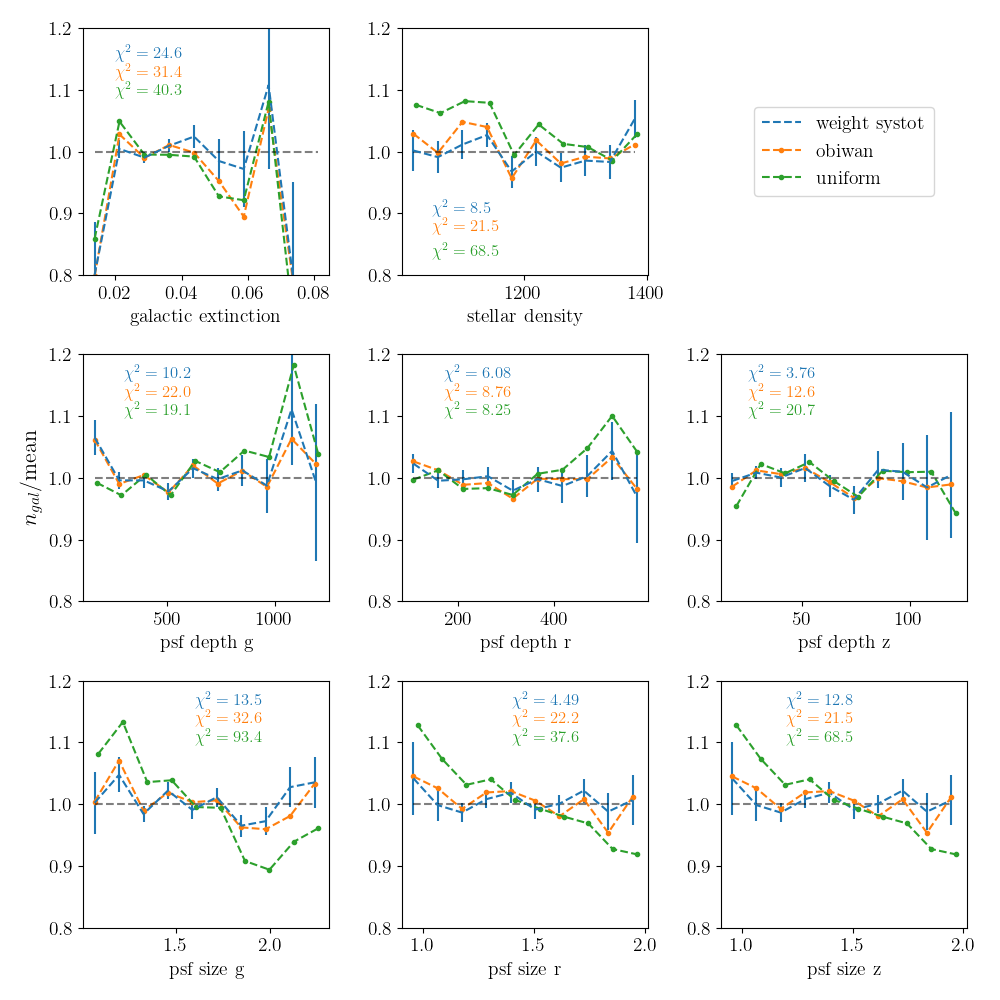}
\caption{The normalized number density at different photometric property bins for eBOSS ELGs for three treatments of the data: 1) applying the weights used in the full eBOSS ELG sample (`weight systot', which is referred to as $w_{\rm sys}$ within the main text; blue;  \citealt{raichoor20a}); 2) applying weights determined from \obiwan\, (`obiwan'; orange); and 3) applying no weights (`uniform', blue). The weights for $w_{\rm sys}$ were determined directly regressing against these maps. The $\chi^2$ test quantifies how the total distribution deviates from the black dashed reference line.}
\label{fig:imaging-systematics}
\end{center}
\end{figure*}

Fig. \ref{fig:imaging-systematics} displays the raw results when using the ELGs in chunk 23 and the uniform randoms (green), when using \obiwan\, randoms (orange), and when applying $w_{\rm{sys}}$ and using uniform randoms (blue with error-bars). From the $\chi^2$ test, we can see that \obiwan \, successfully predicted the degree of correlation with many maps. \obiwan \, is able to handle the complexities for imaging systematics like the correlation between different properties. For example, there is a correlation between stellar density and galactic extinction because they both trace the structure of the Milky Way. \obiwan\, naturally includes all correlations between the maps. In many cases, e.g.,  the mean size of the point spread function of across each CCD image contributing to a given location (psf size), these maps have quantities that we hypothesize could affect our ability to extract sources from the images. \obiwan\, naturally represents a superset of all possible maps and removes the need to identify and classify multiple effects on source detection and measurement. 

The results when applying $w_{\rm{sys}}$ naturally produce the lowest $\chi^2$ values, as all of these maps were used in the regression that produced the $w_{\rm{sys}}$ values.
Using \obiwan\, represents a simpler, and more complete, method for modeling systematic effects related to the detection and measurement of photometric properties of galaxies, and it gives an alternative result for explaining how systematics work in the process. However, the correlation with psf size is still strong when applying \obiwan\,. There is a negative trend with increasing psf size and a $\chi^2$/dof greater than 20/10 for each band. Currently, \obiwan\, uses the PSF file generated by {\legacypipe} during its calibration stage as a truth PSF input, and it is thus not able to trace any imperfections in PSF modeling stage. This indicates that the initial PSF modeling stage may not be perfect. This will affect \obiwan\, twice, as \obiwan\, uses PSF both for modeling input galaxies and also fitting for output galaxies. For the true measurements, only the 2nd stage is effected. Despite this imperfection, {\obiwan} is clearly able to capture most of the effect of the PSF size, as the $\chi^2$ improves significantly in each band, compared to the uniform random case.

\subsection{The Correlation Function}
\label{sec:results-cf}

To compute the correlation function with and without \obiwan, we select galaxies and randoms from the eBOSS ELG catalogs (described in \citealt{raichoor20a}), restricted to chunk 23. We compare results using $w_{\rm sys}$, no imaging systematic weight, and a weight constructed from the \obiwan\, outputs.

We create the \obiwan\, weights as follows: We use a {\textsc healpix} map at $N_{\rm side}=$ 128 resolution and count the number of \obiwan-ELGs in each pixel. We divide this by the number of uniform randoms in each pixel and then normalize by the overall ratio of uniform randoms to \obiwan-ELGs. The inverse of this can then be used as a weight applied to each eBOSS ELG in the same way as $w_{\rm sys}$. We denote this as `\obiwan-weight'. We then substitute \obiwan-weight for $w_{\rm sys}$ when determining $\xi_{0,2,4}$. This provides a simpler comparison than if we were to attempt to directly use the \obiwan\, outputs as randoms for the calculation of $\xi$.

The $w_{\rm sys}$ weights are calculated by directly regressing the ELG data against the template maps. Thus, the method will artificially remove true clustering modes that align with the templates by chance. This will slightly depress the clustering. \obiwan\, is immune from this effect. We predict the size of the effect using GLAM-QPM mocks generated for the eBOSS ELG sample as described in \cite{lin20a}.
 We apply the regression to each mock to provide each with $w_{\rm sys}$, without putting in any systematic trends, and calculate $\xi_{0,2,4}$ with and without $w_{\rm sys}$. When comparing the eBOSS ELG $\xi$ results between using \obiwan\, and $w_{\rm sys}$, we subtract the difference we find in the GLAM-QPM mocks from the \obiwan\, result in order to provide a fair comparison.\footnote{The conclusions we present would be unchanged if we ignore this correction, but having calculated it, we apply it.}

\begin{figure}
     \label{fig:w-theta}{
       \includegraphics[width=0.8\columnwidth]{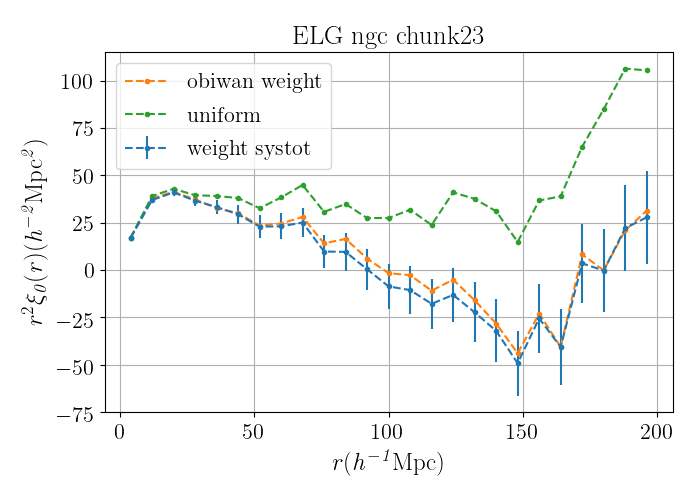}
     }
     \hfill
     \label{fig:thetaw-theta}{
       \includegraphics[width=0.8\columnwidth]{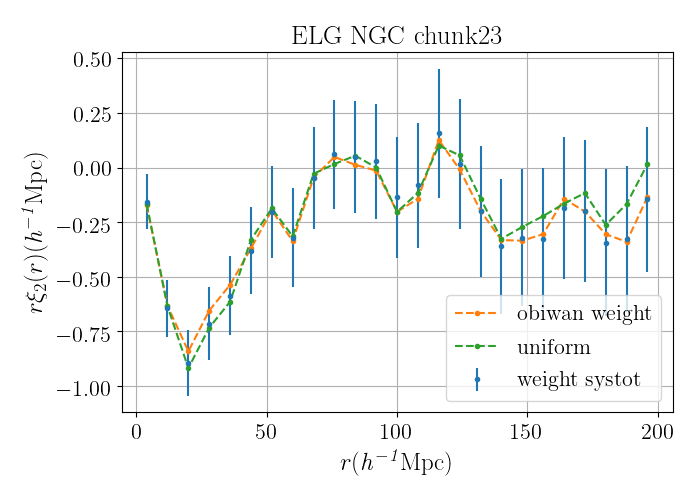}
     }
     \hfill
     \label{fig:thetaw-theta-zoom}{
       \includegraphics[width=0.8\columnwidth]{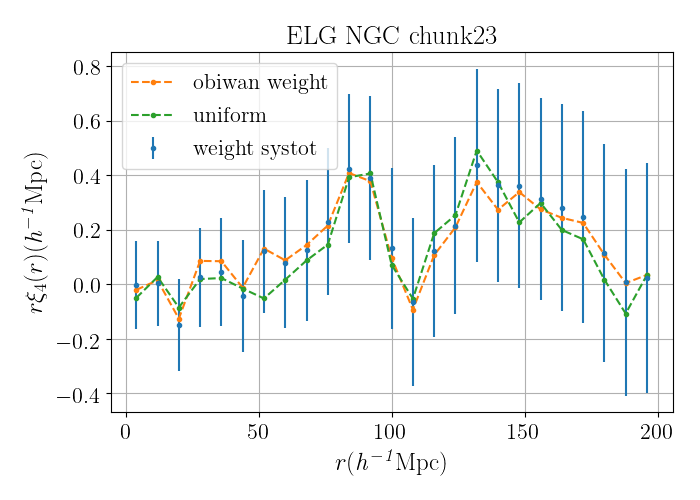}
     }
\caption{Correlation functions of monopole (upper), quadrapole (middle), and hexadecapole (lower), applying weights as in the same three cases described in Fig. \ref{fig:imaging-systematics}.  We use 25 evenly spaced $r$ bins, centered between 0 and 200 $h^{-1}$Mpc.}
\label{fig:cf}
\end{figure}

Fig. \ref{fig:cf} compares the multipoles of the redshift-space correlation function of eBOSS ELGs for three cases: 1) \obiwan-weights (the new method proposed in this paper); 2) no imaging systematic weights (denoted `uniform'); 3) using the $w_{\rm sys}$ weights included in the eBOSS ELG catalogs. The \obiwan-weights correlation function agrees well with the ELG correlation function generated using the $w_{\rm sys}$ weights (orange) \cite{raichoor20a}, while the uniform-randoms monopole shows a considerable excess, especially at large-scales. The quadrupole and hexadecapole display only minor differences between the three cases. The error-bars in these figures are the diagonal elements of the covariance matrix constructed using the eBOSS ELG EZmocks, as described in Section \ref{sec:methods-cf}.

To investigate the significance of the difference between the \obiwan\, and $w_{\rm sys}$ results, we calculate the $\chi^2$ between the two cases for each multipole. Taking the square-root of this value can be considered as the `maximum possible difference' in terms of standard deviations in calculating cosmological parameters (i.e., if some parameter happened to align perfectly with the difference in the correlation functions). We determine the $\chi^2$ values for the measurements with $r>$ 24Mpc$\cdot h^{-1}$, as scales smaller than this are not used in the cosmological analysis \citep{raichoor20a}. For the monopole, $\chi^2$=1.26, despite the fact that one can observe noticeably reduced clustering amplitudes in the range $75 < r < 150$Mpc$\cdot h^{-1}$ for $w_{\rm sys}$ compared to \obiwan-weight. For the quadrupole and hexadecapole, we find differences are even smaller; $\chi^2=0.54$ and 0.31, respectively, and no features are apparent in the figures. 

We do not find any contamination in the eBOSS ELG chunk 23 clustering that is not already removed by $w_{\rm sys}$. This is important as \obiwan\, is not available in the other three eBOSS chunks. Our results suggests that any systematic issues that remain in the ELG catalogs would not be discovered by \obiwan. The main noticeable difference between the \obiwan\, and $w_{\rm sys}$ results is a slight excess in the \obiwan\, monopole. The difference is possibly because of the residual dependence we find with PSF size when using \obiwan. We plan to improve the modeling of the PSF in future \obiwan\, work.

\section{Conclusions}
\label{sec:conclusions}

We have used the Legacy Survey \obiwan\, image simulation tool \citep{obiwanMethods} to
develop a method to remove the effects of imaging systematics from 3D clustering measurements. The idea is similar to \cite{balrog}, with key developments to allow efficient application to the 3D clustering of redshift survey data. The method is intended for application to the next-generation redshift survey DESI. We take advantage of the fact that the eBOSS ELG sample was targeted with Legacy Survey imaging data in order to test our method. Our method is applied to one of the four chunks of the eBOSS ELG sample and we compare our results to the eBOSS ELG analysis that is applied over all four chunks. Our results can be summarized as:

\begin{enumerate}
\item We inject potential ELGs into the images and then determine those that end up passing the eBOSS ELG sample color/magnitude cuts. This allowed us to determine the \gb, \rband, \zb--mag distributions for recovered ELGs (true ELGs that remain ELGs after source detection and measurement), contaminants (non--ELGs that pass target selection after detection and measurement), and lost ELGs (ELGs that are not detected, fail target selection after measurement, or overlap CCD edges). We also investigated how much scattering occurs ($\sim 0.25$ mag) into and out of the eBOSS ELG color box and \gb--band limit. See Figs. \ref{fig:rec-lost-contam-mag} and \ref{fig:rec-lost-contam-color}.

\item Redshifts are applied to our injected potential ELGs from a mixture of eBOSS, VVDS, and DEEP2 data, such that the recovered distribution is a good match to the eBOSS distribution (see Section \ref{sec:injecting-elgs}). Thus, we can also estimate the redshift distributions for recovered ELGs, contaminants, etc. Fig. \ref{fig:redshifts-recovered} shows that the contaminant redshifts are primarily $z<0.75$. This is consistent with the findings that the eBOSS ELG dN/dz depends strongly on the imaging depth at $z< 0.75$ \citep{raichoor20a,demattia20a}.

\item The angular fluctuations in recovered \obiwan\ ELGs mimic those imparted by image properties that can be mapped. Fig. \ref{fig:imaging-systematics} demonstrates that after correcting for the density predicted by \obiwan\, the relative eBOSS ELG density vs. imaging properties is close to the null expectation. A small residual trend is left with the PSF size, implying this is an area where further improvements to the \obiwan\, methods could be made.

\item We applied the angular selection function predicted by \obiwan\ as a weight to correct for imaging systematics. We compared 3D clustering results using this weight to those obtained from the weights (based on linear regression) used in the full eBOSS ELG sample cosmological analysis. We find minor differences, plausibly consistent with the difference in the residuals against PSF size. See Fig. \ref{fig:cf}. The differences are such that, given similar results over the full eBOSS ELG footprint, we expect minimal impact on the cosmological results. This is thus an important validation of the methods applied to the full eBOSS ELG sample.

\item Finally, we identified numerous biases and systematics in the DR5 Legacy Surveys image reduction pipeline, \legacypipe. The highest impact ones are that \legacypipe\, underestimates the uncertainty on \gb, \rband, and \zb\, flux by a factor of 1.7--1.9, the uncertainty on $\rhalf$ by a factor of 2.3--3.9, and the uncertainty on e1 and e2 by a factor of 4.1--4.4. Many of these issues have been fixed in the DR8 version of the pipeline. See Appendix \ref{sec:biases-systematics} for more details. 
\end{enumerate}

\section*{Data Availability}
The eBOSS ELG catalog data and the corresponding \obiwan\ outputs will be made available through the SDSS Science Archive Server (https://sas.sdss.org/) after this work is accepted for publication, with the exact location to be determined.

\section*{Acknowledgements} 
\label{sec:ack}

Funding for the DEEP2 Galaxy Redshift Survey has been provided by NSF grants AST-95-09298, AST-0071048, AST-0507428, and AST-0507483 as well as NASA LTSA grant NNG04GC89G.

This research uses data from the VIMOS VLT Deep Survey, obtained from the VVDS database operated by Cesam, Laboratoire d'Astrophysique de Marseille, France.

Funding for the Sloan Digital Sky Survey IV has been provided by the Alfred P. Sloan Foundation, the U.S. Department of Energy Office of Science, and the Participating Institutions. SDSS-IV acknowledges
support and resources from the Center for High-Performance Computing at
the University of Utah. The SDSS web site is www.sdss.org.

SDSS-IV is managed by the Astrophysical Research Consortium for the 
Participating Institutions of the SDSS Collaboration including the 
Brazilian Participation Group, the Carnegie Institution for Science, 
Carnegie Mellon University, the Chilean Participation Group, the French Participation Group, Harvard-Smithsonian Center for Astrophysics, 
Instituto de Astrof\'isica de Canarias, The Johns Hopkins University, Kavli Institute for the Physics and Mathematics of the Universe (IPMU) / 
University of Tokyo, the Korean Participation Group, Lawrence Berkeley National Laboratory, 
Leibniz Institut f\"ur Astrophysik Potsdam (AIP),  
Max-Planck-Institut f\"ur Astronomie (MPIA Heidelberg), 
Max-Planck-Institut f\"ur Astrophysik (MPA Garching), 
Max-Planck-Institut f\"ur Extraterrestrische Physik (MPE), 
National Astronomical Observatories of China, New Mexico State University, 
New York University, University of Notre Dame, 
Observat\'ario Nacional / MCTI, The Ohio State University, 
Pennsylvania State University, Shanghai Astronomical Observatory, 
United Kingdom Participation Group,
Universidad Nacional Aut\'onoma de M\'exico, University of Arizona, 
University of Colorado Boulder, University of Oxford, University of Portsmouth, 
University of Utah, University of Virginia, University of Washington, University of Wisconsin, 
Vanderbilt University, and Yale University.

JM gratefully acknowledges support from the U.S. Department of Energy,
Office of Science, Office of High Energy Physics under Award Number
DE-SC002008 and from the National Science Foundation under grant
AST-1616414.


\bibliographystyle{mnras}
\bibliography{bib}

\begin{thebibliography}{}
\makeatletter
\relax
\def\mn@urlcharsother{\let\do\@makeother \do\$\do\&\do\#\do\^\do\_\do\%\do\~}
\def\mn@doi{\begingroup\mn@urlcharsother \@ifnextchar [ {\mn@doi@}
  {\mn@doi@[]}}
\def\mn@doi@[#1]#2{\def\@tempa{#1}\ifx\@tempa\@empty \href
  {http://dx.doi.org/#2} {doi:#2}\else \href {http://dx.doi.org/#2} {#1}\fi
  \endgroup}
\def\mn@eprint#1#2{\mn@eprint@#1:#2::\@nil}
\def\mn@eprint@arXiv#1{\href {http://arxiv.org/abs/#1} {{\tt arXiv:#1}}}
\def\mn@eprint@dblp#1{\href {http://dblp.uni-trier.de/rec/bibtex/#1.xml}
  {dblp:#1}}
\def\mn@eprint@#1:#2:#3:#4\@nil{\def\@tempa {#1}\def\@tempb {#2}\def\@tempc
  {#3}\ifx \@tempc \@empty \let \@tempc \@tempb \let \@tempb \@tempa \fi \ifx
  \@tempb \@empty \def\@tempb {arXiv}\fi \@ifundefined
  {mn@eprint@\@tempb}{\@tempb:\@tempc}{\expandafter \expandafter \csname
  mn@eprint@\@tempb\endcsname \expandafter{\@tempc}}}

\bibitem[\protect\citeauthoryear{Abbott et~al.,}{Abbott
  et~al.}{2019}]{abbott2019dark}
Abbott T.,  et~al., 2019, Physical Review D, 99, 123505

\bibitem[\protect\citeauthoryear{Ahumada et~al.,}{Ahumada
  et~al.}{2019}]{ahumada2019sixteenth}
Ahumada R.,  et~al., 2019, The Sixteenth Data Release of the Sloan Digital Sky
  Surveys: First Release from the APOGEE-2 Southern Survey and Full Release of
  eBOSS Spectra (\mn@eprint {arXiv} {1912.02905})

\bibitem[\protect\citeauthoryear{{Alam} et~al.}{{Alam} et~al.}{2020}]{Alam20}
{Alam} S.,  et~al., 2020, submitted

\bibitem[\protect\citeauthoryear{{Avila} et~al.}{{Avila}
  et~al.}{2020}]{Avila20}
{Avila} S.,  et~al., 2020, submitted

\bibitem[\protect\citeauthoryear{{Bautista} et~al.}{{Bautista}
  et~al.}{2020}]{LRG_corr}
{Bautista} J.,  et~al., 2020, submitted

\bibitem[\protect\citeauthoryear{{Blake} et~al.,}{{Blake}
  et~al.}{2010}]{wigglezSelectionFunc}
{Blake} C.,  et~al., 2010, \mn@doi [\mnras] {10.1111/j.1365-2966.2010.16747.x},
  \href {http://adsabs.harvard.edu/abs/2010MNRAS.406..803B} {406, 803}

\bibitem[\protect\citeauthoryear{{Blanton} et~al.,}{{Blanton}
  et~al.}{2017}]{sdss4}
{Blanton} M.~R.,  et~al., 2017, \mn@doi [\aj] {10.3847/1538-3881/aa7567}, \href
  {https://ui.adsabs.harvard.edu/abs/2017AJ....154...28B} {154, 28}

\bibitem[\protect\citeauthoryear{{Burleigh}}{{Burleigh}}{2018}]{obiwanMethods}
{Burleigh} K.~J.,  2018, PhD thesis, UC berkeley

\bibitem[\protect\citeauthoryear{Burleigh et~al.,}{Burleigh
  et~al.}{2020}]{burleigh2020observing}
Burleigh K.~J.,  et~al., 2020, Observing Strategy for the Legacy Surveys
  (\mn@eprint {arXiv} {2002.05828})

\bibitem[\protect\citeauthoryear{Chuang, Kitaura, Prada, Zhao  \& Yepes}{Chuang
  et~al.}{2014}]{10.1093/mnras/stu2301}
Chuang C.-H.,  Kitaura F.-S.,  Prada F.,  Zhao C.,   Yepes G.,  2014, \mn@doi
  [Monthly Notices of the Royal Astronomical Society] {10.1093/mnras/stu2301},
  446, 2621

\bibitem[\protect\citeauthoryear{{Collaboration} et~al.}{{Collaboration}
  et~al.}{2020}]{eBOSS_Cosmology}
{Collaboration} e.,  et~al., 2020, submitted

\bibitem[\protect\citeauthoryear{{Colless} et~al.,}{{Colless}
  et~al.}{2001}]{2dFGRS}
{Colless} M.,  et~al., 2001, \mn@doi [\mnras]
  {10.1046/j.1365-8711.2001.04902.x}, \href
  {http://adsabs.harvard.edu/abs/2001MNRAS.328.1039C} {328, 1039}

\bibitem[\protect\citeauthoryear{{DESI Collaboration} et~al.,}{{DESI
  Collaboration} et~al.}{2016a}]{desiScience}
{DESI Collaboration} et~al., 2016a, preprint, \href
  {http://adsabs.harvard.edu/abs/2016arXiv161100036D} {} (\mn@eprint {arXiv}
  {1611.00036})

\bibitem[\protect\citeauthoryear{{DESI Collaboration} et~al.,}{{DESI
  Collaboration} et~al.}{2016b}]{desiInstrument}
{DESI Collaboration} et~al., 2016b, preprint, \href
  {http://adsabs.harvard.edu/abs/2016arXiv161100037D} {} (\mn@eprint {arXiv}
  {1611.00037})

\bibitem[\protect\citeauthoryear{{Dawson} et~al.,}{{Dawson}
  et~al.}{2013}]{bossSurvey}
{Dawson} K.~S.,  et~al., 2013, \mn@doi [\aj] {10.1088/0004-6256/145/1/10},
  \href {http://adsabs.harvard.edu/abs/2013AJ....145...10D} {145, 10}

\bibitem[\protect\citeauthoryear{{Dawson} et~al.,}{{Dawson}
  et~al.}{2016}]{dawson2016sdss}
{Dawson} K.~S.,  et~al., 2016, \mn@doi [\aj] {10.3847/0004-6256/151/2/44},
  \href {https://ui.adsabs.harvard.edu/abs/2016AJ....151...44D} {151, 44}

\bibitem[\protect\citeauthoryear{{Delubac} et~al.,}{{Delubac}
  et~al.}{2017}]{delubacSystematics}
{Delubac} T.,  et~al., 2017, \mn@doi [\mnras] {10.1093/mnras/stw2741}, \href
  {http://adsabs.harvard.edu/abs/2017MNRAS.465.1831D} {465, 1831}

\bibitem[\protect\citeauthoryear{{Dey} et~al.,}{{Dey}
  et~al.}{2018}]{overviewPaper}
{Dey} A.,  et~al., 2018, preprint, \href
  {https://ui.adsabs.harvard.edu/#abs/2018arXiv180408657D} {p.
  arXiv:1804.08657} (\mn@eprint {arXiv} {1804.08657})

\bibitem[\protect\citeauthoryear{{Drinkwater} et~al.,}{{Drinkwater}
  et~al.}{2010}]{wigglezSurvey}
{Drinkwater} M.~J.,  et~al., 2010, \mn@doi [\mnras]
  {10.1111/j.1365-2966.2009.15754.x}, \href
  {http://adsabs.harvard.edu/abs/2010MNRAS.401.1429D} {401, 1429}

\bibitem[\protect\citeauthoryear{{Elsner}, {Leistedt}  \& {Peiris}}{{Elsner}
  et~al.}{2016}]{biasInTemplateMethod}
{Elsner} F.,  {Leistedt} B.,   {Peiris} H.~V.,  2016, \mn@doi [\mnras]
  {10.1093/mnras/stv2777}, \href
  {http://adsabs.harvard.edu/abs/2016MNRAS.456.2095E} {456, 2095}

\bibitem[\protect\citeauthoryear{{Elvin-Poole} et~al.,}{{Elvin-Poole}
  et~al.}{2017}]{elvinpoole}
{Elvin-Poole} J.,  et~al., 2017, preprint, \href
  {http://adsabs.harvard.edu/abs/2017arXiv170801536E} {} (\mn@eprint {arXiv}
  {1708.01536})

\bibitem[\protect\citeauthoryear{{Falco} et~al.,}{{Falco}
  et~al.}{1999}]{cfaTwo}
{Falco} E.~E.,  et~al., 1999, \mn@doi [\pasp] {10.1086/316343}, \href
  {http://adsabs.harvard.edu/abs/1999PASP..111..438F} {111, 438}

\bibitem[\protect\citeauthoryear{{Favole} et~al.,}{{Favole}
  et~al.}{2016}]{corrfuncEboss}
{Favole} G.,  et~al., 2016, \mn@doi [\mnras] {10.1093/mnras/stw1483}, \href
  {https://ui.adsabs.harvard.edu/#abs/2016MNRAS.461.3421F} {461, 3421}

\bibitem[\protect\citeauthoryear{Fevre et~al.,}{Fevre
  et~al.}{2014}]{fevre:hal-01113687}
Fevre O.~L.,  et~al., 2014, {Astronomy and Astrophysics - A\&A}, pp 559, 14

\bibitem[\protect\citeauthoryear{{Flaugher} et~al.,}{{Flaugher}
  et~al.}{2015}]{DECam}
{Flaugher} B.,  et~al., 2015, \mn@doi [\aj] {10.1088/0004-6256/150/5/150},
  \href {https://ui.adsabs.harvard.edu/abs/2015AJ....150..150F} {150, 150}

\bibitem[\protect\citeauthoryear{{Gil-Marin} et~al.}{{Gil-Marin}
  et~al.}{2020}]{gil-marin20a}
{Gil-Marin} H.,  et~al., 2020, submitted

\bibitem[\protect\citeauthoryear{{G{\'o}rski}, {Hivon}, {Banday}, {Wandelt},
  {Hansen}, {Reinecke}  \& {Bartelmann}}{{G{\'o}rski} et~al.}{2005}]{healpix}
{G{\'o}rski} K.~M.,  {Hivon} E.,  {Banday} A.~J.,  {Wandelt} B.~D.,  {Hansen}
  F.~K.,  {Reinecke} M.,   {Bartelmann} M.,  2005, \mn@doi [\apj]
  {10.1086/427976}, \href {http://adsabs.harvard.edu/abs/2005ApJ...622..759G}
  {622, 759}

\bibitem[\protect\citeauthoryear{{Gunn} et~al.,}{{Gunn} et~al.}{2006}]{Gunn06}
{Gunn} J.~E.,  et~al., 2006, \mn@doi [\aj] {10.1086/500975}, \href
  {https://ui.adsabs.harvard.edu/abs/2006AJ....131.2332G} {131, 2332}

\bibitem[\protect\citeauthoryear{{Hamilton}}{{Hamilton}}{1993}]{corrfuncHamilton}
{Hamilton} A.~J.~S.,  1993, \mn@doi [\apj] {10.1086/173288}, \href
  {https://ui.adsabs.harvard.edu/#abs/1993ApJ...417...19H} {417, 19}

\bibitem[\protect\citeauthoryear{{Ho} et~al.,}{{Ho}
  et~al.}{2012}]{sdss8Companion}
{Ho} S.,  et~al., 2012, \mn@doi [\apj] {10.1088/0004-637X/761/1/14}, \href
  {http://adsabs.harvard.edu/abs/2012ApJ...761...14H} {761, 14}

\bibitem[\protect\citeauthoryear{{Hou} et~al.}{{Hou} et~al.}{2020}]{hou20a}
{Hou} J.,  et~al., 2020, submitted

\bibitem[\protect\citeauthoryear{{Huchra}, {Vogeley}  \& {Geller}}{{Huchra}
  et~al.}{1999}]{cfaOne}
{Huchra} J.~P.,  {Vogeley} M.~S.,   {Geller} M.~J.,  1999, \mn@doi [\apjs]
  {10.1086/313194}, \href {http://adsabs.harvard.edu/abs/1999ApJS..121..287H}
  {121, 287}

\bibitem[\protect\citeauthoryear{{Kaiser}}{{Kaiser}}{1987}]{Kaiser87}
{Kaiser} N.,  1987, \mn@doi [\mnras] {10.1093/mnras/227.1.1}, \href
  {https://ui.adsabs.harvard.edu/abs/1987MNRAS.227....1K} {227, 1}

\bibitem[\protect\citeauthoryear{{Landy} \& {Szalay}}{{Landy} \&
  {Szalay}}{1993}]{landy93}
{Landy} S.~D.,  {Szalay} A.~S.,  1993, \mn@doi [\apj] {10.1086/172900}, \href
  {http://adsabs.harvard.edu/abs/1993ApJ...412...64L} {412, 64}

\bibitem[\protect\citeauthoryear{{Lang} et~al.,}{{Lang}
  et~al.}{prep}]{tractorPaper}
{Lang} D.,  et~al., in prep., \aj

\bibitem[\protect\citeauthoryear{{Laurent} et~al.,}{{Laurent}
  et~al.}{2017}]{qsoDepthExtinction}
{Laurent} P.,  et~al., 2017, \mn@doi [\jcap] {10.1088/1475-7516/2017/07/017},
  \href {http://adsabs.harvard.edu/abs/2017JCAP...07..017L} {7, 017}

\bibitem[\protect\citeauthoryear{{Le F{\`e}vre} et~al.,}{{Le F{\`e}vre}
  et~al.}{2005}]{2005A&A...439..845L}
{Le F{\`e}vre} O.,  et~al., 2005, \mn@doi [\aap] {10.1051/0004-6361:20041960},
  \href {https://ui.adsabs.harvard.edu/abs/2005A%26A...439..845L} {439, 845}

\bibitem[\protect\citeauthoryear{{Leistedt}, {Peiris}, {Mortlock},
  {Benoit-L{\'e}vy}  \& {Pontzen}}{{Leistedt} et~al.}{2013}]{leistedt13}
{Leistedt} B.,  {Peiris} H.~V.,  {Mortlock} D.~J.,  {Benoit-L{\'e}vy} A.,
  {Pontzen} A.,  2013, \mn@doi [\mnras] {10.1093/mnras/stt1359}, \href
  {http://adsabs.harvard.edu/abs/2013MNRAS.435.1857L} {435, 1857}

\bibitem[\protect\citeauthoryear{{Lin} et~al.}{{Lin} et~al.}{2020}]{lin20a}
{Lin} S.,  et~al., 2020, submitted

\bibitem[\protect\citeauthoryear{{Myers} et~al.,}{{Myers}
  et~al.}{2006a}]{sdss1Systematics}
{Myers} A.~D.,  et~al., 2006a, \mn@doi [\apj] {10.1086/499093}, \href
  {http://adsabs.harvard.edu/abs/2006ApJ...638..622M} {638, 622}

\bibitem[\protect\citeauthoryear{{Myers} et~al.,}{{Myers}
  et~al.}{2006b}]{myers06}
{Myers} A.~D.,  et~al., 2006b, \mn@doi [\apj] {10.1086/499093}, \href
  {http://adsabs.harvard.edu/abs/2006ApJ...638..622M} {638, 622}

\bibitem[\protect\citeauthoryear{{Myers} et~al.,}{{Myers}
  et~al.}{2015}]{myersRegressionTech}
{Myers} A.~D.,  et~al., 2015, \mn@doi [\apjs] {10.1088/0067-0049/221/2/27},
  \href {http://adsabs.harvard.edu/abs/2015ApJS..221...27M} {221, 27}

\bibitem[\protect\citeauthoryear{{Neveux} et~al.}{{Neveux}
  et~al.}{2020}]{neveux20a}
{Neveux} R.,  et~al., 2020, submitted

\bibitem[\protect\citeauthoryear{{Newman} et~al.,}{{Newman}
  et~al.}{2013}]{deep2}
{Newman} J.~A.,  et~al., 2013, \mn@doi [\apjs] {10.1088/0067-0049/208/1/5},
  \href {http://adsabs.harvard.edu/abs/2013ApJS..208....5N} {208, 5}

\bibitem[\protect\citeauthoryear{{Norberg}, {Baugh}, {Gazta{\~n}aga}  \&
  {Croton}}{{Norberg} et~al.}{2009}]{corrfuncErrors}
{Norberg} P.,  {Baugh} C.~M.,  {Gazta{\~n}aga} E.,   {Croton} D.~J.,  2009,
  \mn@doi [\mnras] {10.1111/j.1365-2966.2009.14389.x}, \href
  {https://ui.adsabs.harvard.edu/#abs/2009MNRAS.396...19N} {396, 19}

\bibitem[\protect\citeauthoryear{{Peebles}}{{Peebles}}{1980}]{peebles1980}
{Peebles} P.~J.~E.,  1980, {The large-scale structure of the universe}

\bibitem[\protect\citeauthoryear{{Prakash} et~al.,}{{Prakash}
  et~al.}{2016}]{prakashRegressionTech}
{Prakash} A.,  et~al., 2016, \mn@doi [\apjs] {10.3847/0067-0049/224/2/34},
  \href {http://adsabs.harvard.edu/abs/2016ApJS..224...34P} {224, 34}

\bibitem[\protect\citeauthoryear{{Raichoor} et~al.,}{{Raichoor}
  et~al.}{2017}]{anand17}
{Raichoor} A.,  et~al., 2017, \mn@doi [\mnras] {10.1093/mnras/stx1790}, \href
  {http://adsabs.harvard.edu/abs/2017MNRAS.471.3955R} {471, 3955}

\bibitem[\protect\citeauthoryear{{Raichoor} et~al.}{{Raichoor}
  et~al.}{2020}]{raichoor20a}
{Raichoor} A.,  et~al., 2020, submitted

\bibitem[\protect\citeauthoryear{{Ross} et~al.,}{{Ross}
  et~al.}{2011}]{sdss8Systematics}
{Ross} A.~J.,  et~al., 2011, \mn@doi [\mnras]
  {10.1111/j.1365-2966.2011.19351.x}, \href
  {http://adsabs.harvard.edu/abs/2011MNRAS.417.1350R} {417, 1350}

\bibitem[\protect\citeauthoryear{{Ross} et~al.,}{{Ross}
  et~al.}{2012}]{sdss9Systematics}
{Ross} A.~J.,  et~al., 2012, \mn@doi [\mnras]
  {10.1111/j.1365-2966.2012.21235.x}, \href
  {http://adsabs.harvard.edu/abs/2012MNRAS.424..564R} {424, 564}

\bibitem[\protect\citeauthoryear{{Ross} et~al.,}{{Ross}
  et~al.}{2017}]{sdss12Systematics}
{Ross} A.~J.,  et~al., 2017, \mn@doi [\mnras] {10.1093/mnras/stw2372}, \href
  {http://adsabs.harvard.edu/abs/2017MNRAS.464.1168R} {464, 1168}

\bibitem[\protect\citeauthoryear{{Ross} et~al.}{{Ross} et~al.}{2020}]{ross20a}
{Ross} A.~J.,  et~al., 2020, submitted

\bibitem[\protect\citeauthoryear{{Rossi} et~al.}{{Rossi}
  et~al.}{2020}]{rossi20a}
{Rossi} G.,  et~al., 2020, submitted

\bibitem[\protect\citeauthoryear{{Rybicki} \& {Press}}{{Rybicki} \&
  {Press}}{1992}]{rybicki92}
{Rybicki} G.~B.,  {Press} W.~H.,  1992, \mn@doi [\apj] {10.1086/171845}, \href
  {https://ui.adsabs.harvard.edu/#abs/1992ApJ...398..169R} {398, 169}

\bibitem[\protect\citeauthoryear{{Sawangwit}, {Shanks}, {Abdalla}, {Cannon},
  {Croom}, {Edge}, {Ross}  \& {Wake}}{{Sawangwit} et~al.}{2011}]{corrfuncIC}
{Sawangwit} U.,  {Shanks} T.,  {Abdalla} F.~B.,  {Cannon} R.~D.,  {Croom}
  S.~M.,  {Edge} A.~C.,  {Ross} N.~P.,   {Wake} D.~A.,  2011, \mn@doi [\mnras]
  {10.1111/j.1365-2966.2011.19251.x}, \href
  {https://ui.adsabs.harvard.edu/#abs/2011MNRAS.416.3033S} {416, 3033}

\bibitem[\protect\citeauthoryear{{Slosar}, {Seljak}  \& {Makarov}}{{Slosar}
  et~al.}{2004}]{uros04}
{Slosar} A.,  {Seljak} U.,   {Makarov} A.,  2004, \mn@doi [\prd]
  {10.1103/PhysRevD.69.123003}, \href
  {http://adsabs.harvard.edu/abs/2004PhRvD..69l3003S} {69, 123003}

\bibitem[\protect\citeauthoryear{{Smee} et~al.,}{{Smee} et~al.}{2013}]{Smee13}
{Smee} S.~A.,  et~al., 2013, \mn@doi [\aj] {10.1088/0004-6256/146/2/32}, \href
  {https://ui.adsabs.harvard.edu/abs/2013AJ....146...32S} {146, 32}

\bibitem[\protect\citeauthoryear{{Smith} et~al.}{{Smith}
  et~al.}{2020}]{smith20}
{Smith} A.,  et~al., 2020, submitted

\bibitem[\protect\citeauthoryear{{Stetson}}{{Stetson}}{1987}]{1987PASP...99..191S}
{Stetson} P.~B.,  1987, \mn@doi [\pasp] {10.1086/131977}, \href
  {https://ui.adsabs.harvard.edu/abs/1987PASP...99..191S} {99, 191}

\bibitem[\protect\citeauthoryear{{Suchyta} et~al.,}{{Suchyta}
  et~al.}{2016}]{balrog}
{Suchyta} E.,  et~al., 2016, \mn@doi [\mnras] {10.1093/mnras/stv2953}, \href
  {http://adsabs.harvard.edu/abs/2016MNRAS.457..786S} {457, 786}

\bibitem[\protect\citeauthoryear{{Tamone} et~al.}{{Tamone}
  et~al.}{2020}]{tamone20a}
{Tamone} A.,  et~al., 2020, submitted

\bibitem[\protect\citeauthoryear{{Tegmark}, {Hamilton}, {Strauss}, {Vogeley}
  \& {Szalay}}{{Tegmark} et~al.}{1998}]{tegmark98}
{Tegmark} M.,  {Hamilton} A. J.~S.,  {Strauss} M.~A.,  {Vogeley} M.~S.,
  {Szalay} A.~S.,  1998, \mn@doi [\apj] {10.1086/305663}, \href
  {https://ui.adsabs.harvard.edu/#abs/1998ApJ...499..555T} {499, 555}

\bibitem[\protect\citeauthoryear{{Weinberg}, {Mortonson}, {Eisenstein},
  {Hirata}, {Riess}  \& {Rozo}}{{Weinberg} et~al.}{2013}]{weinberg}
{Weinberg} D.~H.,  {Mortonson} M.~J.,  {Eisenstein} D.~J.,  {Hirata} C.,
  {Riess} A.~G.,   {Rozo} E.,  2013, \mn@doi [\physrep]
  {10.1016/j.physrep.2013.05.001}, \href
  {https://ui.adsabs.harvard.edu/#abs/2013PhR...530...87W} {530, 87}

\bibitem[\protect\citeauthoryear{{York} et~al.,}{{York}
  et~al.}{2000}]{sdssYork}
{York} D.~G.,  et~al., 2000, \mn@doi [\aj] {10.1086/301513}, \href
  {http://adsabs.harvard.edu/abs/2000AJ....120.1579Y} {120, 1579}

\bibitem[\protect\citeauthoryear{{Zhao} et~al.}{{Zhao} et~al.}{2020}]{zhao20a}
{Zhao} C.,  et~al., 2020, submitted

\bibitem[\protect\citeauthoryear{{de Mattia} et~al.}{{de Mattia}
  et~al.}{2020}]{demattia20a}
{de Mattia} A.,  et~al., 2020, submitted

\makeatother
\end{thebibliography}



\appendix

\section{Biases and Systematics (\legacypipe)}
\label{sec:biases-systematics}

\begin{figure}
\begin{center}
 \includegraphics[width=0.98\columnwidth]{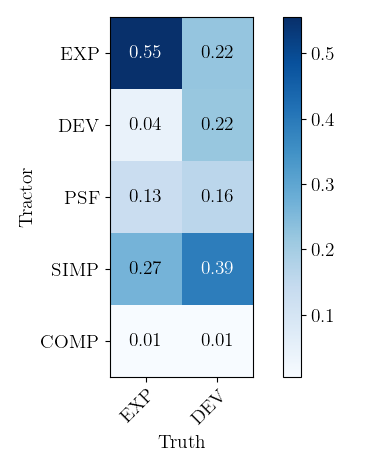}
\end{center}

 \caption{Confusion matrix showing the fraction of true exponential or \dev\, sources that \tractor\, models as type PSF, SIMP, EXP, DEV, or COMP. \tractor\, is biased towards EXP sources because 91\% of true exponential sources are modeled as exponential, while 23\% of true \dev\, sources are modeled as \dev.}
 \label{fig:confusion}
\end{figure}

This section describes the biases and systematics that we find in the DR5 version of \legacypipe\, after running \obiwan\, on eBOSS chunk23 CCDs to reproduce eBOSS ELG data. We expect significant improvement in the DR7 version of \legacypipe\, \citep{overviewPaper}. Fig. \ref{fig:confusion} shows that \tractor\, is biased towards EXP sources. About 95\% of true exponential sources are modeled as exponential (PSF, SIMP are both special cases of exponential galaxies), while 20\% of true \dev\, sources are modeled as \dev. The other 80\% of truly \dev\, sources are classified as SIMP (45\%), EXP (25\%), and PSF (7\%). The EXP bias is surprising because \tractor\, model selection penalizes EXP and DEV sources equally (see \citealt{obiwanMethods}).

\begin{figure}
\begin{center}
 \includegraphics[width=0.96\columnwidth]{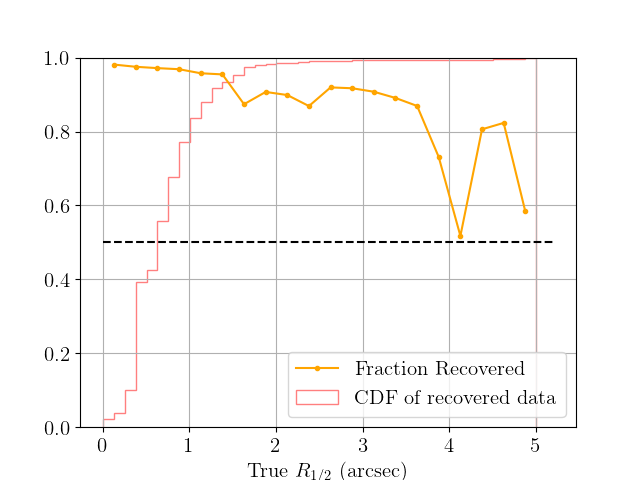}
\end{center}

 \caption{(yellow) Fraction of all sources recovered by \legacypipe\, versus injected $\rhalf$. (red) Cumulative histrogram of injected $\rhalf$.}
 \label{fig:rhalf-input}
\end{figure}

Fig. \ref{fig:rhalf-input} shows the fraction of all sources recovered by \legacypipe\, versus true $\rhalf$. There is a characteristic size ($\rhalf \sim 1.5$\arcsec) after which the fraction recovered drops to, and fluctuates about, 90\%. This does not have a significant impact on the final data sample because 90\% the input galaxies are within this characteristic size ($\rhalf \sim 1.5$\arcsec).

\begin{figure*}
\begin{center}
 \includegraphics[width=1.9\columnwidth]{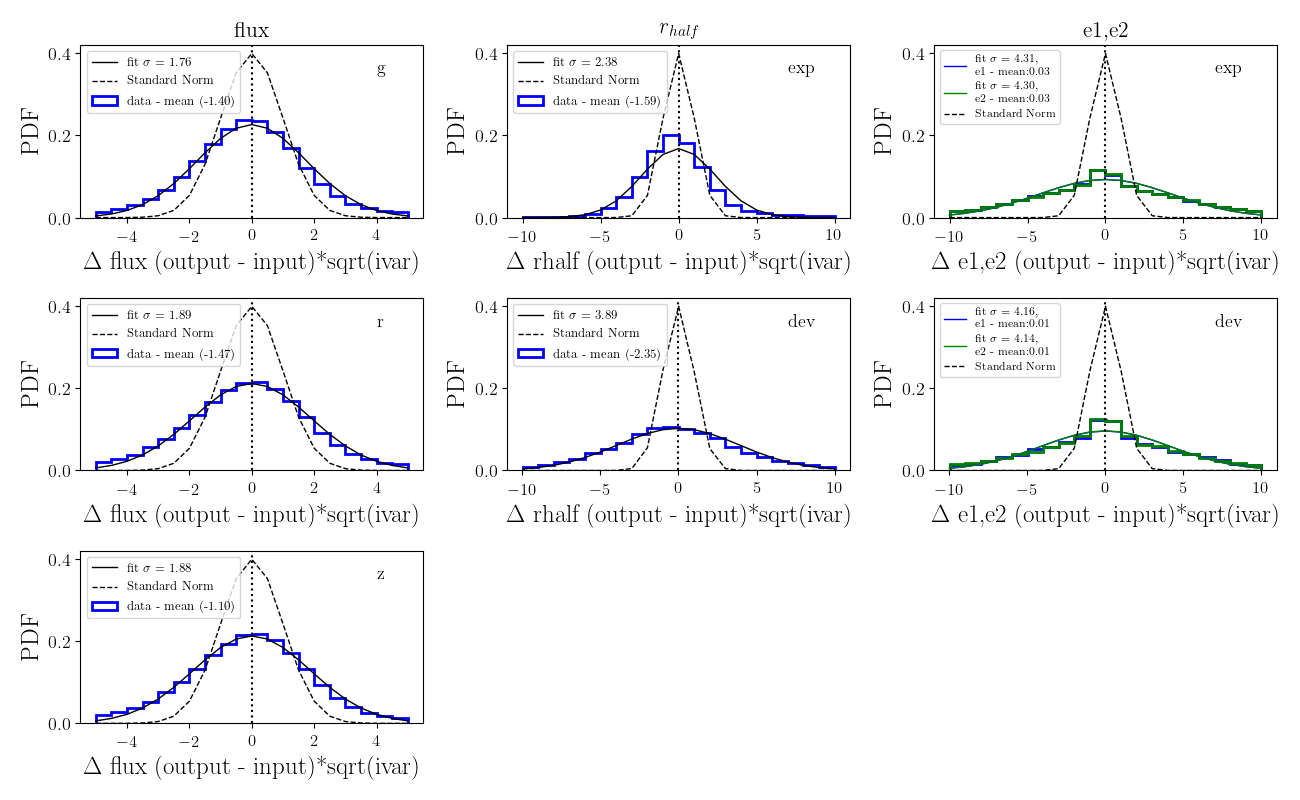}
\end{center}
\caption{Number of standard deviations away from truth (input; $N_\sigma$) of the \legacypipe\, measured flux, $\rhalf$, and ellipticity. The mean of each distribution has been subtracted. The least squares fit Gaussians (solid black) are considerably wider than a Normal Gaussian (dashed black). This suggests that all of \legacypipe\, measurement errors are underestimated. (Left) $N_\sigma$  for \gb, \rband, \zb\, flux. (Middle) $N_\sigma$ for $\rhalf$ for sources \tractor\, classifies as EXP and DEV. (Right) $N_\sigma$ for ellipticity e1 and e2 for sources \tractor\, classifies as EXP and DEV. 
\label{fig:tractor-uncertainty}}
\end{figure*}

Fig. \ref{fig:tractor-uncertainty} shows the number of standard deviations ($N_\sigma$) away from truth of the \tractor\, measured \gb, \rband, and \zb--band flux, $\rhalf$, and ellipticity e1 and e2. There are very large systematic offset in flux ($\sim 0.25$ mag in all bands) and $\rhalf$ ($\sim $1.5--2.4\arcsec\, for EXP and DEV sources), which we remove by subtracting the mean. \tractor\, fluxes are too faint while \tractor\, $\rhalf$ is too large. There is no systematic offset for the ellipticity e1 and e2 measurements. The least squares fit Gaussians (solid black) are considerably wider than a Normal Gaussian (dashed black). This suggests that all of \tractor\, measurement errors are underestimated, by factors of 1.75--1.9x for \gb, \rband, \zb\, flux, 2.3--3,9x for $\rhalf$, and 4.1--4.3x for ellipticity e1 and e2.

\begin{figure*}
\begin{center}
 \includegraphics[width=1.9\columnwidth]{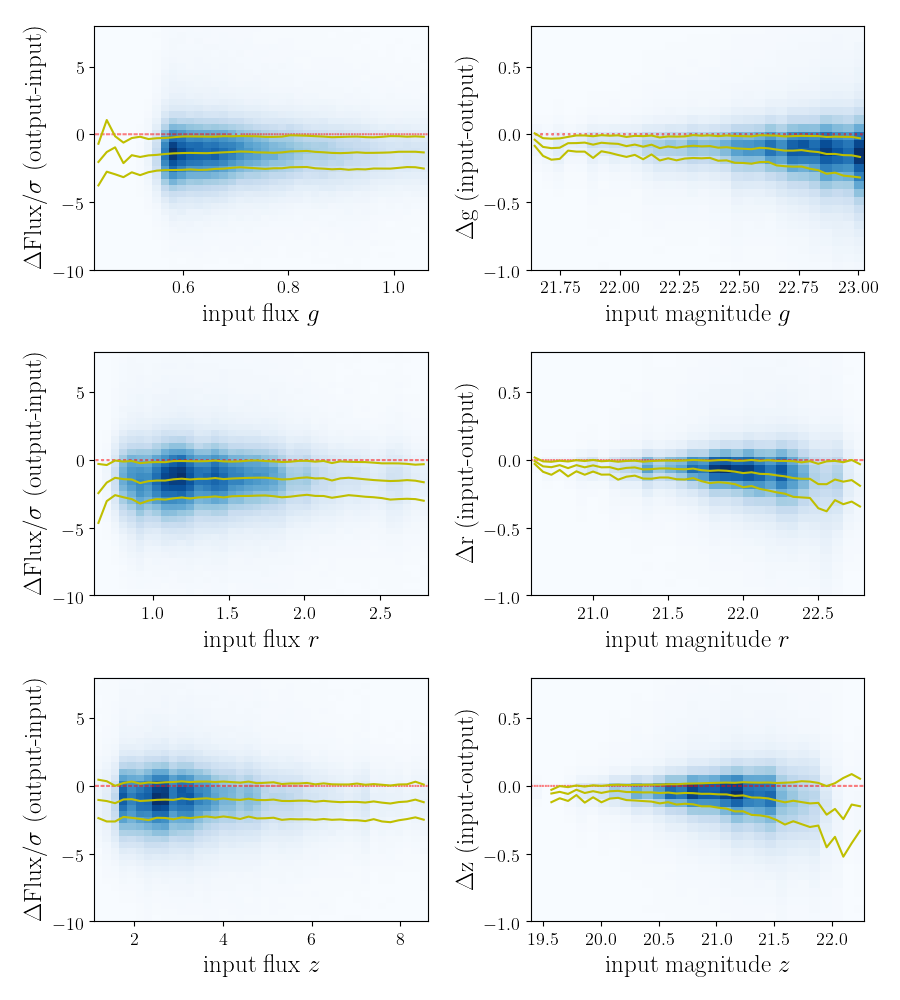}
\caption{2--dimensional histograms of \obiwan-input -\-- \obiwan-output residuals. (Left) $N_\sigma$ for \gb, \rband, \zb\, flux versus \gb, \rband, \zb\, magnitude, respectively. Yellow lines are the q25, 50, 75 percentiles. (Right) Magnitude residuals versus  \gb, \rband, \zb\, magnitude.}
\label{fig:num-std-dev-and-dmag}
\end{center}
\end{figure*}

The measurement uncertainty is naturally dependent on the level of signal. Thus, we constructed 2--dimensional histograms of $N_\sigma$ for \gb, \rband, \zb\, flux versus \gb, \rband, \zb\, flux, respectively (Fig. \ref{fig:num-std-dev-and-dmag}, left panel); and magnitude residual versus \gb, \rband, \zb\, magnitude (Fig. \ref{fig:num-std-dev-and-dmag}, right panel). The results are nearly identical when only considering PSF, SIMP, EXP, or DEV sources. We find the bias in the recovered magnitude does increase as sources grow fainter but that it is nearly constant as a 1$\sigma$ bias in the flux.

\section{Injecting Realistic eBOSS ELGs}

\subsection{ELG-like Targets}
\label{sec:eboss-almost-targets}

\begin{figure}
\begin{center}
 \includegraphics[width=0.99\columnwidth]{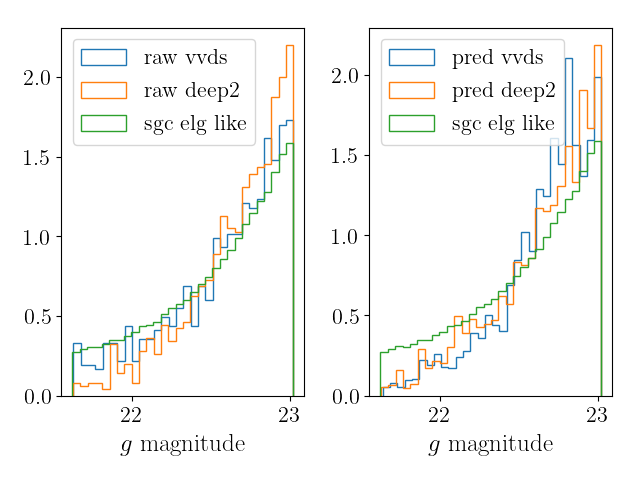}
\end{center}
 \caption{(left) $g$ magnitude distribution of DR3--VVDS sample (blue), DR3--DEEP2 sample (orange) and eBOSS ELG--like sample. (right) 50000 draws from redshift sampling (described in Section \ref{sec:injecting-elgs}) by DR3--VVDS sample (blue), DR3--DEEP2 sample (orange). eBOSS ELG--like sample (green) is for reference. }
 \label{fig:dr3dp2-reproduces}
\end{figure}

\begin{figure}
\begin{center}

\includegraphics[width=0.95\columnwidth]{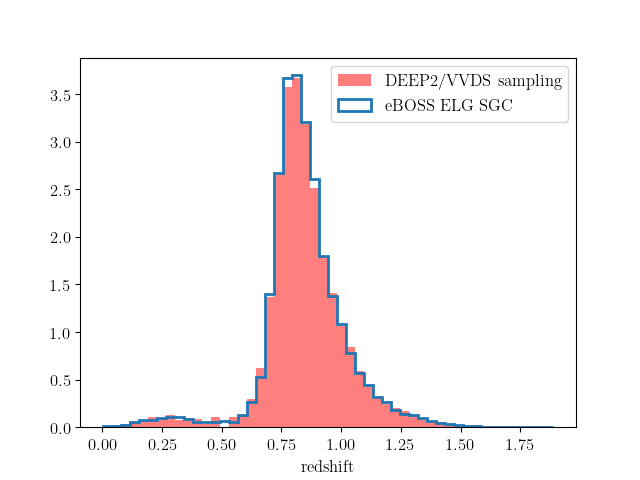}

\end{center}
\caption{The blue curve displays the normalized histogram of eBOSS ELG redshifts in the SGC region. The filled-red region is the same for 50,000 draws from the Gaussian Mixture Model used to sample redshifts fro from the DEEP2/VVDS data.}
\label{fig:nz}
\end{figure}

As stated in section \ref{sec:elg-like}, we use ELG-like targets as our inputs. Fig.~\ref{fig:dr3dp2-reproduces} (left) shows the $g$ magnitude distribution of DR3--DEEP2 sample and DR3--VVDS sample, as well as the ELG--like sample in SGC. Before the redshift sampling (discussed in \ref{sec:injecting-elgs}), the $g$ magnitude distribution of DR3--VVDS looks more like eBOSS ELG--like galaxies than DR3--DEEP2 sample. However, after the resampling, Fig. \ref{fig:dr3dp2-reproduces} (right) shows that VVDS and DEEP2 sample looks quite similar. This is because redshift sampling modifies redshift distribution, and thus magnitude distribution also gets modified. We force them to have the same redshift distribution. This redshift distribution is sampled from using a a 10-component Gaussian Mixture Model (GMM) fit to the ELG redshift distribution in the SGC. This distribution is shown in Fig. \ref{fig:nz}.

\begin{figure}
\begin{center}
\includegraphics[width=\columnwidth]{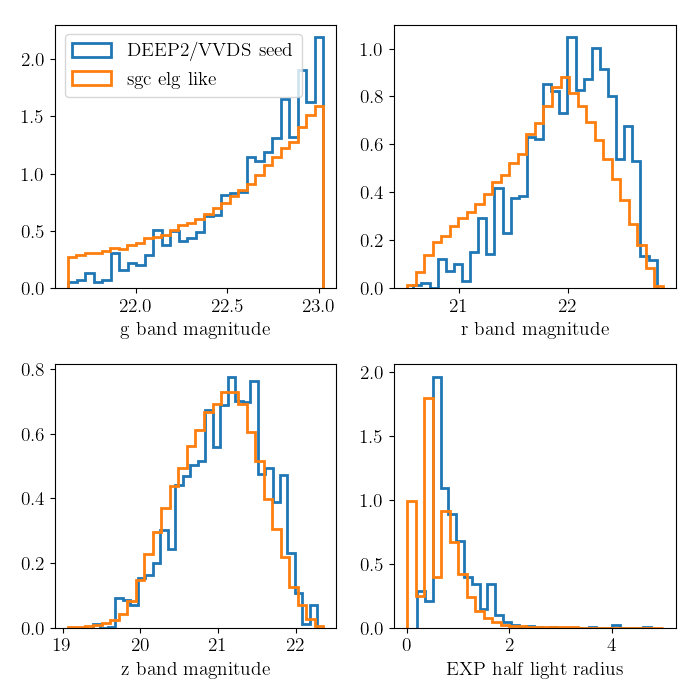}
\end{center}

 \caption{The color and half light radius distribution of Exponential galaxies for two samples: the DR3-DEEP2-VVDS combined sample (DEEP2/VVDS seed; blue) and the DR3 photometric sample cut to the eBOSS ELG-like color box (sgc elg like; orange).}
 \label{fig:final-eboss-sample-10k}
\end{figure}

The SGC ELG-like sample is different from the DR3-DEEP2 and DR3-VVDS samples. This is because we do not know the redshift distribution of all the SGC ELG-like sample, and its true redshift distribution is different from the eBOSS ELG sample. We combine DR3-DEEP2 sample and DR3-VVDS sample to form our final DR3-DEEP2-VVDS sample of 1849 EXP galaxies and 150 DEV galaxies. Fig. \ref{fig:final-eboss-sample-10k} shows the 50000 draws from DR3--DEEP2--VVDS sampled galaxies. They do not have the same distribution as the SGC ELG-like sample. This is due to the fact that we have weighted the DEEP2 and VVDS galaxies such that the eBOSS $dN/dz$ is reproduced. Fig. \ref{fig:dr3dp2-vs-eboss} shows that after cutting these 50000 draws based on the \obiwan\, outputs and the eBOSS ELG color/magnitude selection, the color, magnitude, and redshift distributions are a close match to the eBOSS ELG ones.

\bsp
\label{lastpage}
\end{document}